\documentclass[12pt]{article}
\newlength{\abstractwidth}
\pdfoutput=1

\usepackage{amsmath}
\usepackage{epsf}
\usepackage{color}
\usepackage{graphicx}
\usepackage{amsfonts, amssymb}

\renewcommand{\thefootnote}{\fnsymbol{footnote}}
\renewcommand{\thanks}[1]{\footnote{#1}}
\newcommand{\starttext}{
\setcounter{footnote}{0}
\renewcommand{\thefootnote}{\arabic{footnote}}}

\newcommand{\bea}{\begin{eqnarray}}
\newcommand{\eea}{\end{eqnarray}}
\newcommand{\ee}{\end{equation}}
\newcommand{\be}{\begin{equation}}

\begin{document}
\starttext
\setcounter{footnote}{0}
\begin{flushright}
\today
\end{flushright}

\bigskip

\begin{center}

{\LARGE \bf Domain Wall Holography} \\
 \medskip
{\LARGE \bf for Finite Temperature Scaling Solutions}

\vskip .4in

{\large \bf Eric Perlmutter\footnote{ericperlmutter@gmail.com}}

\vskip .2in

{ \sl Department of Physics and Astronomy }\\
{\sl University of California, Los Angeles, CA 90095, USA}\\

\end{center}

\vskip .2in

\begin{abstract}

\vskip 0.1in 

We investigate a class of near-extremal solutions of Einstein-Maxwell-scalar theory with electric charge and power law scaling, dual to charged IR phases of relativistic field theories at low temperature. These are exact solutions of theories with domain wall vacua; hence, we use nonconformal holography to relate the bulk and boundary theories. We numerically construct a global interpolating solution between the IR charged solutions and the UV domain wall vacua for arbitrary physical choices of Lagrangian parameters. By passing to a conformal frame in which the domain wall metric becomes that of AdS, we uncover a generalized scale invariance of the IR scaling solution, indicating a connection to the physics of Lifshitz fixed points. Finally, guided by effective field theoretic principles and the physics of nonconformal D-branes, we argue for the applicability of domain wall holography even in theories with AdS critical points, namely those theories for which a scalar potential is dominated by a single exponential term over a large range.

\end{abstract}

\newpage

\contentsline {section}{\numberline {1}Introduction and summary}{2}
\contentsline {section}{\numberline {2}Construction of the action and scaling solution}{7}
\contentsline {subsection}{\numberline {2.1}Extremal}{10}
\contentsline {subsection}{\numberline {2.2}Near-extremal}{12}
\contentsline {section}{\numberline {3}Domain walls}{13}
\contentsline {subsection}{\numberline {3.1}Geometry}{14}
\contentsline {subsection}{\numberline {3.2}Holography}{17}
\contentsline {section}{\numberline {4}Numerical construction of interpolating solution}{21}
\contentsline {section}{\numberline {5}Generalized scale invariance}{25}
\contentsline {section}{\numberline {6}Domain wall holography and effective field theory}{28}
\contentsline {section}{\numberline {7}Discussion and prospects}{32}
\contentsline {section}{\numberline {A}Numerical study of finite temperature modified Lifshitz solution}{33}

\section{Introduction and summary}
\setcounter{equation}{0}

The form and function of the AdS/CFT correspondence \cite{Maldacena:1997re} have evolved since the idea's birth. Initially, the correspondence was borne from consideration of the decoupling limit of branes in string and M-theory with conformal near-horizon supersymmetries, including the paradigmatic case of duality between type IIB supergravity on AdS$_5\times S^5$ and strongly coupled $\mathcal{N}=4$ super-Yang-Mills theory in the large N limit. Subsequent work relaxed the correspondence to include bulk spacetimes that are asymptotically AdS and to those with non-maximal or no supersymmetry.\footnote{For foundational work and a review, see \cite{gkp,witten,magoo}.} The last few years have also seen a broadening of the scope of correspondence to include bulk actions that have no origin in string or M-theory: one simply writes down a ``bottom-up" Lagrangian that admits an AdS vacuum, and asks what sort of field theories it can describe, while tabling questions regarding the quantum existence of the bulk theory. This has led to a productive interface between AdS/CFT and condensed matter physics, including constructions of gravity duals to holographic superconductors \cite{hhh,Gubserstringc,Gauntlett}, non-relativistic theories with anisotropic Lifshitz scaling \cite{mulligan,taylor,pal}, and theories with Galilean invariance \cite{Balasubramanian,son}, among much other work.

There has also been work on theories with no microscopic scaling symmetry at all, but rather merely relativistic symmetry. This correspondence \cite{mald2} models strongly coupled, Poincar\'e-invariant quantum field theories (possibly with some supersymmetry) with a bulk spacetime that is, at least asymptotically, of the form
\begin{equation}\label{DW1}
ds^2=(A r)^{\gamma '} \eta_{\mu\nu}dx^{\mu}dx^{\nu} + \frac{dr^2}{(A r)^{\gamma '}} \, ,
\end{equation}
along with some number of rolling scalar fields. This goes by the name ``Domain Wall/QFT correspondence"\cite{townsend}.\footnote{Henceforth labeled ``DW/QFT" for short.} The fact that the metric (\ref{DW1}) is conformal to AdS can be phrased as crucial to the duality, which was established by decoupling the near-horizon dynamics of non-conformal D-branes: passage to a conformal, so-called ``dual" frame in which the D-brane near-horizon metric becomes AdS$_{p+2}\times S^{8-p}$ reveals a manifest $(p+2)$-dimensional gravitational description, in which we identify the radial direction with the energy scale of the dual field theory \cite{polchinskipeet}. Because the scalars vary and the curvature is somewhere singular, the duality holds at intermediate energies, away from large curvature and order one effective string coupling. 

This picture can be summed up by the statement that these nonconformal D-branes have a ``generalized conformal structure"\cite{skenderis,Jevicki:1998ub}: upon passage to the dual frame, the conformal symmetry of the metric is broken by the nonconstant scalar field, implying that the radial scale transformation leaves the solution invariant if one shifts the scalar field simultaneously. This follows from the conformal structure of M-theory branes and their relation to ten-dimensional type II branes.

In \cite{skenderis,wisemanwithers} a precise holographic dictionary was established in this case, both for choices of $\gamma '$ that derive from branes and those that do not -- that is, the program of holographic renormalization was extended to these non-asymptotically AdS spacetimes. It seems natural then to extend the phenomenological philosophy described above in pursuit of the question, ``what are all IR spacetimes that can be patched onto a domain wall solution?" This dualizes to the question, ``what long-range behavior is possible for systems with merely Poincar\'e symmetry?" This approach was utilized in \cite{skenderis2} in the context of hydrodynamics of nonconformal branes, for example, but we would like to continue to study phases of field theories to which DW/QFT applies.

In a separate development, we recall that gravity duals of condensed matter systems often have the undesirable feature of nonzero entropy at extremality. This runs counter to the empirical Nernst's ``Theorem", which says that a generic physical system cooled to zero temperature should lose all its entropy as it occupies a unique ground state. Gauge/gravity duality translates this to the statement that we should study finite temperature black holes which have a degenerating horizon as the temperature is tuned to zero, unlike the standard example of the charged Reissner-Nordstrom black brane.\footnote{Of course, this near-horizon AdS$_2$ region of the Reissner-Nordstrom black hole has also been used constructively, for instance in modeling behavior of IR CFTs in 0+1 dimensions. See e.g. \cite{mcgreevyvegh1,Faulkner:2009wj,polchfaulk,Faulkner:2010da}.} Much work has succeeded in understanding both the nature of low-temperature physics in strongly coupled systems, and the extent to which gravity duals can model the intricacies of the real world. But we would like to understand this better; for instance, intuition might suggest that such ground state entropies disappear when generic interactions are turned on, or when introducing a nonzero external field, but these notions have yet to be comprehensively confirmed or rebutted (see e.g. \cite{per1,per2}).

In this paper, we show that these two puzzles -- modeling systems with Poincar\'e symmetry and understanding the universal aspects of strongly coupled condensed matter systems -- find novel intersection. Specifically, we consider finite temperature scaling solutions of a theory of gravity coupled to a U(1) gauge field and a real, neutral scalar, with a scalar-dependent gauge coupling. In this general form, the action resembles various dimensionally-reduced string theory actions. We make an electric ansatz for the gauge field, so our theory will describe a field theory with finite electric charge density. Explicitly, we examine an action
\begin{equation}\label{actionintro}
\begin{split} S = -\frac{1}{16\pi G_D} \int d^{D}x \sqrt{-g} \Big(R&+f(\phi)F_{\mu\nu}F^{\mu\nu} +\frac{1}{2}(\partial\phi)^2 + \mathcal{V}(\phi)\Big) \, , \\ 
\end{split}
\end{equation}
which admits a general scaling solution for the metric,
\begin{equation}\label{scaling1}
ds^2\sim -r^{\beta}f(r)dt^2+\frac{dr^2}{r^{\beta}f(r)}+ r^{\gamma} dx^i\cdot dx^i
\end{equation}
where $f(r)$ is a near-extremal ``emblackening" factor that vanishes at the near-extremal horizon, $r=r_h$, and equals one in the extremal solution ($r_h=0$). The field equations demand $\beta, \gamma >0$, so this metric, by design, has zero extremal entropy. Hence, its extremal limit can be viewed as approximating the leading near-horizon behavior of a solution with vanishing horizon area; that is, as the far infrared gravity dual to a system with a unique zero temperature ground state. 

If one insists that this class of metrics is an exact solution of the theory (\ref{actionintro}), it can only be supported with rolling scalars of the form 
\begin{equation}\label{rollingscalar}
\phi(r) \sim \ln{r}+\phi_0
\end{equation}
Moreover, the field equations demand that the scalar potential and gauge coupling are single exponentials in $\phi$,
\begin{equation}\label{pot}
\mathcal{V}(\phi)=-V_0 e^{\eta\phi} \, , \quad f(\phi)=e^{\alpha\phi} \, .
\end{equation}

An action of the form (\ref{actionintro}) with this exponential potential clearly does not admit AdS; in fact, its flux-less vacuum is a domain wall. Therefore, if the scaling solution (\ref{scaling1}) is patched onto an asymptotically domain wall spacetime, then DW/QFT holography dualizes the global solution to a relativistic theory with power law thermodynamic scaling at low energies and temperatures. This interpolating geometry -- which we construct numerically -- connects the IR spacetime (\ref{scaling1}) to the UV spacetime (\ref{DW1}), with electric flux sourcing the IR geometry and a logarithmically rolling scalar accompanying both.

Just as the asymptotic domain wall of the global solution is conformal to AdS, one can ask whether applying that same Weyl transformation to the IR scaling solution reveals any similar generalized symmetry left unbroken by the electric charge. Indeed, we find a conformal frame scale-invariance in the IR metric, so that the near-extremal scaling metric becomes the near-extremal Lifshitz solution, 
\begin{equation}\label{lifshitz}
\widetilde{ds}^2_{IR}=-\frac{r^{2}}{l^2}f(r)dt^2+\frac{l^2}{r^{2}}\frac{dr^2}{f(r)}+\left(\frac{r}{l}\right)^{\tilde{\gamma}}dx^i\cdot dx^i \, ,
\end{equation}
modified by a logarithmically rolling scalar (\ref{rollingscalar}). 

The low-energy physics of our dual field theory, therefore, is controlled by a ``generalized scale invariance," despite the solution not exhibiting scale invariance in Einstein frame. Of course, the invariance is broken by the finite temperature, but it is nonetheless an interesting result, natural in the context of DW/QFT: just as the UV domain wall spacetime possesses a generalized conformal structure, the charged, Lorentz-symmetry breaking phase to which the theory flows in the IR retains the part of this structure unbroken by the presence of that charge. 

Four-dimensional holography for the scaling solution (\ref{scaling1}) was studied for the scale-invariant $\beta=2$ case in \cite{kachru}, both numerically and analytically, and extended to arbitrary spacetime dimension in \cite{asian}. We refer to this as the ``modified Lifshitz solution." The five-dimensional $\beta \neq 2$ solution was presented in \cite{gubser}, where the authors studied a charged dilaton AdS$_5$ black hole with a ten-dimensional uplift to spinning D3 branes. That full dilatonic black hole solution approaches this one, for certain values of $\eta$ and $\alpha$, in the limit that the scalar field is large and one can neglect an exponential term in its potential. We noted earlier that the single-power scaling of the metric is characteristic of the leading small $r$ behavior of some solution near an extremal horizon of vanishing area. Restricting our attention to only that term, and demanding that it be part of an exact solution to the action (\ref{actionintro}), we found that the potential must be a single exponential in $\phi$, namely the term in some full potential which dominates near the horizon, of course; this was indeed the motivation of \cite{gubser} to consider the scaling solution at all.

It was also suggested in \cite{gubser} that a deformation of the potential $\mathcal{V}(\phi)$ which generates an AdS critical point would allow one to study the Einstein frame solution (\ref{scaling1}) holographically via AdS/CFT. That would necessarily turn the solution into an IR phase of a conformal, not merely quantum, field theory. We take a different approach, asking as we did earlier about IR behavior of theories with relativistic symmetry instead. 

Even in the conformal case, one should be able to use DW/QFT at intermediate energies below the scale at which conformal symmetry is manifest, where a single term of a full (even stringy) potential dominates. We argue for this role of DW/QFT in the body of the paper, guided by the philosophy of effective field theory. Consider, for instance, a scalar potential of the form
\begin{equation}\label{poten}
\mathcal{V}(\phi)=-V_0(e^{-b\phi}+be^{\phi}) \, , \quad 1 < 2b+1 < \sqrt{1+\frac{(D-1)^2}{V_0}} \, ,
\end{equation}
and $V_0>0$. This admits an AdS vacuum at the origin, stable with respect to the Breitenlohner-Freedman bound. When $\phi$ is large, the potential is dominated by the second term. So for a solution in which $\phi$ goes to $+\infty$ at the horizon, then decreases monotonically outwards and settles at its $\phi=0$ AdS critical point, the fields behave as though near a domain wall boundary over a large range of $r$. This suggests that domain wall holography can act as an effective holographic tool at intermediate energies where a theory is only relativistic, and the terms in the full potential that become large at larger radii (higher energies) are not needed.

The paper is organized as follows. We first show, in section 2, that such scaling behavior for the metric is only compatible with single exponential scalar potentials and gauge coupling functions. In the process, we come to bear on why there is attractor behavior in the extremal modified Lifshitz case, despite the presence of a scalar field that breaks the isometry: the functional form of the scalar is fixed once the metric is given. We construct the scaling solution in terms of a fixed Lagrangian and explore its parameter space. Section 3 reviews the relevant aspects of domain wall geometries and introduces the DW/QFT correspondence itself. We argue that only a subclass of all domain walls can be treated holographically, namely those with a boundary. The meat of the paper begins in section 4, where we describe and show evidence of the numerical construction of the global solution, which is of scaling form in the IR and domain wall form in the UV. We put this to use in section 5, where we expose a generalized scale invariance of the scaling solution that descends from the generalized conformal structure of the domain wall. In section 6, we argue for the wider applicability of domain wall holography, and classify when it is safe to use, by drawing on effective field theory principles and lessons from the case of nonconformal D-branes. Section 7 concludes with a discussion of the results and prospects for future work. 

\medskip
\textit{Note:} As this work was being finalized, two papers appeared which give different treatments of the same solutions. 

The authors of \cite{charmousis} characterize a wide class of theories that includes ours and provide a thorough analytic investigation. They make the assumption that the theories have AdS critical points in the UV, which we do not, as discussed earlier.  Additionally, we provide the numerical construction of the global solution which explicitly permits holographic analysis. 

The paper \cite{Lee:2010qs} complements our work by calculating conductivities of the scaling solution.


\section{Construction of the action and scaling solution}
\setcounter{equation}{0}
We reproduce the action of our D-dimensional Einstein-Maxwell-scalar bulk system as
\begin{equation}
\begin{split} S = -\frac{1}{16\pi G_D} \int d^{D}x \sqrt{-g} \Big(R&+f(\phi)F_{\mu\nu}F^{\mu\nu} +\frac{1}{2}(\partial\phi)^2+ \mathcal{V}(\phi)\Big) \, , \\ 
\end{split}
\end{equation}
with no reference to any string or M-theory origin as yet.\footnote{Throughout the paper, we work in $D \geq 4$, avoiding the peculiarities of lower-dimensional gravity.}  $f(\phi)$ is a positive definite function, to ensure the correct sign for the gauge kinetic term. We will consider solutions with electric charge only, so we do not write any Chern-Simons terms; we also do not show the boundary terms required for the usual construction of a well-defined variational problem. 

Our electric, non-relativistic, planar-symmetric ansatz is
\begin{equation}
\begin{split}
&A_t=A_t(r) \, , \,  A_r=\vec{A} = 0\,  \\ 
&\phi=\phi(r) \, \\
&ds^{2} = -U(r)dt^{2}+\frac{dr^{2}}{U(r)}+V(r)dx^i\cdot dx^i \, ,\\
\end{split}
\end{equation} 
where $i$ indexes the $D-2$ boundary spatial coordinates. This metric describes planar black holes, putatively dual to an electrically charged boundary theory living in Minkowski space.

Our first goal is to show that the scaling behavior for the metric discussed earlier is only compatible with an exponential gauge coupling function and scalar potential, and a logarithmically rolling scalar field. In the case of a constant potential, we recover AdS$_D$, AdS$_2\times\mathbb{R}^{D-2}$, and the modified Lifshitz geometry as limiting cases of the most general solution. 

Extracting the field equations, the $t$ component of Maxwell's equations can be integrated to give the field strength,
\begin{equation}
F_{rt}=A_t'(r)=\frac{\rho}{f(\phi)V^{\frac{D-2}{2}}} \, ;
\end{equation}
the integration constant $\rho$ acts as the charge density of the black hole. Writing the rest of the field equations in these terms, we have the scalar equation,
\begin{equation}
S:\quad U\phi '' + \Big(\frac{D-2}{2}\Big) \frac{UV'}{V}\phi' + U'\phi ' = \frac{d \mathcal{V}(\phi)}{d\phi}-\frac{2\rho^2}{f^{2}(\phi)V^{D-2}}\frac{df(\phi)}{d\phi}
\end{equation}
as well as three Einstein equations, 
\begin{equation}
\begin{split}
E1:& \quad \Big(\frac{D-2}{2}\Big)\Big(\frac{V''}{V}-\frac{1}{2}\frac{V'^2}{V^2}\Big) = -\frac{(\phi')^2}{2} \\
CON:& \quad \Big(\frac{D-2}{2}\Big)\frac{U'V'}{V}+\Big(\frac{(D-2)(D-3)}{4}\Big)\frac{UV'^2}{V^2}-\frac{U(\phi')^2}{2} +\mathcal{V}(\phi)+\frac{2\rho^2}{f(\phi)V^{D-2}} = 0 \\
\Big\lbrace E2:& \quad \frac{D-2}{2}U''+\Big(\frac{D-2}{2}\Big)^{2}\frac{U'V'}{V} +\mathcal{V}(\phi)-\frac{2(D-3)\rho^2}{f(\phi)V^{D-2}} = 0 \Big\rbrace 
\end{split}
\end{equation}
The bracketed Einstein equation, $(E2)$, is implied by the rest of the field equations, shown for example by differentiation of the constraint equation $(CON)$ and substitution from the others.

As we plug in the scaling behavior 
\begin{equation}\label{scaling2}
U(r) \sim r^{\beta} \, , \, V(r) \sim r^{\gamma} \, ,
\end{equation}
we note that will only consider solutions with $\beta > 1$, so that our solutions obey the usual definition of extremality, namely $T=r_h=0$, ensuring smooth connection to the finite temperature solutions. 

We also point out that the field equations dictate that $\beta \leq 2$ -- where scale invariance of the metric obtains when $\beta=2$ -- and $0 \leq\gamma\leq 2$. Upon fixing the form of the Lagrangian consistent with admission of the scaling solution, these bounds become clear; so let us proceed.

Returning to the field equations, then, the first Einstein equation (E1) tells us that the scalar field must take the form
\begin{equation}
\phi(r)=C_2 \ln{r} + \phi_0 \, .
\end{equation}
The remaining undetermined functions are those of the scalar field, $f(\phi)$ and $\mathcal{V}(\phi)$. But for a neutral scalar, we can form a linear combination of the Einstein equations in which the scalar only appears in $f(\phi)$: taking $(CON)-(E2)+U\cdot(E1)$, we have

\begin{equation}
\begin{split}
& \frac{U'V'}{V}\Big(\frac{D-2}{2}-\Big(\frac{D-2}{2}\Big)^2\Big) + \frac{UV'^2}{V^2}\Big(\frac{(D-2)(D-4)}{4}\Big) \\ 
& + \Big(\frac{UV''}{V}-U''\Big)\frac{D-2}{2} - \frac{2\rho^2(2-D)}{f(\phi)V^{D-2}} = 0
\end{split}
\end{equation}
The gauge coupling function $f(\phi)$ is fully determined by the metric and therefore, by the first-order equation $(CON)$, so is $\mathcal{V}(\phi)$. Specifically, they are both constrained to be exponential in $\phi$, that is, power law in $r$: plugging in the scaling form of the metric reveals, up to positive constants,
\begin{equation}\label{CON-E2}
f(\phi)\sim\frac{\rho^2}{F(\beta,\gamma)}r^{2-\beta-\gamma(D-2)} \, ,
\end{equation}

where we have defined 
\begin{equation}\label{F}
F(\beta,\gamma)= \Big(\frac{D-4}{2}\Big)\gamma\Big(\beta-\gamma\Big)
 +\beta(\beta-1)-\gamma(\gamma-1) \, .
\end{equation}
Demanding reality of the flux, $\rho^2 \geq 0$, implies $F(\beta,\gamma)>0$: allowed combinations of $\beta$ and $\gamma$  are bounded by the lines $\gamma = \beta$ and $\gamma=\frac{2}{D-2}(1-\beta)$. For $\beta > 1$, the metric must have
\begin{equation}
\beta \geq \gamma \, ,
\end{equation}
where saturation occurs for vanishing flux, e.g. in AdS$_D$ where $\beta=\gamma=2$. This is the cousin of the fact that, for example, the Lifshitz geometry (\ref{lifshitz}) sourced by real two- and three-form fluxes, as in \cite{mulligan}, can only act as a gravity dual to Lifshitz fixed points with $z>1$. Such a similarity in the causal structure of our scaling solution to the Lifshitz solution is our first hint that the two may have some connection.

Furthermore, by plugging this form for $f(\phi)$ back into the Einstein equations, one sees that the potential is also a power law in $r$:
\begin{equation}
\mathcal{V}(\phi)=-\rho^2G(\beta,\gamma)r^{\beta-2} \, ,
\end{equation}
where 
\begin{equation}
	G(\beta,\gamma) \equiv 2\Big(1+\frac{(D-2)\gamma}{F(\beta,\gamma)}\Big(\gamma\Big(\frac{D-2}{2}\Big)+(\beta-1)\Big)\Big) \, .
	\end{equation}\\
$\beta, \gamma > 0$ implies $G(\beta,\gamma)>0$ for $D>3$; thus, $\mathcal{V}(\phi)$ must be negative. And because $\beta \leq 2$, $\mathcal{V}(\phi)$ must diverge at small $r$ or be constant everywhere. As expected, a scale-invariant solution can only solve a theory with the latter: as we rescale $r$, $\phi$ picks up a constant, which must not affect the energy of the theory because we are simply executing a symmetry transformation. 

What we have shown, in the end, is that the scalar rolls down the exponential potential as it nears the horizon, presumably signaling the dive toward zero entropy at zero temperature: $\phi$ is ``looking" for a critical point of the potential, as exists for the unique finite entropy extremal AdS$_2\times\mathbb{R}^{D-2}$ geometry \cite{reall}, but cannot find it. 

Even in the case where the potential $\mathcal{V}(\phi)$ is constant, the exponentiality of the gauge coupling function $f(\phi)$ explains the zero extremal entropy in terms of the attractor mechanism \cite{kachru}: the diverging scalar drives the system toward the runaway minimum of the effective attractor potential, $V_{eff} =\rho^2 f^{-1}(\phi)$. The fact that there is attractor behavior at all despite the lack of true SO(2,1) isometry is accounted for by our analysis above: the form of the attractor potential is fixed once the metric's SO(2,1) isometry is given. Thus we have a case of an attractor in which the full functional form of the massless scalar is fixed near the extremal horizon. 

Let us note that if the scalar is charged, the form of $f(\phi)$ and $\mathcal{V}(\phi)$ is not fixed as above, indicating that a charged interaction between $\phi$ and $A_{\mu}$ is compatible with extremal scaling behavior (\ref{scaling1}) for a range of gauge couplings and potentials.

In anticipation of a possible embedding of this solution into a consistent truncation of some higher-dimensional supergravity, we end this subsection with the observation that a multi-scalar version of this solution can also support the metric (\ref{scaling2}). Writing a schematic action 
\begin{equation}\label{multiaction}
\begin{split} S = -\frac{1}{16\pi G_D} \int d^{D}x \sqrt{-g} \Big(R&+f(\phi _i)F_{\mu\nu}F^{\mu\nu} +\frac{1}{2}(\partial\phi_i)^2+ \mathcal{V}(\phi_i)\Big) \, , \\ 
\end{split}
\end{equation}
field equation (E1) is satisfied for all scalars logarithmic in $r$. Then the power law behavior of $f(\phi_i)$ and $\mathcal{V}(\phi_i)$ means that both are products of exponentials,
\begin{equation}
\mathcal{V}(\phi)=-V_0 e^{\eta_i\phi_i} \, , \quad f(\phi)=e^{\alpha_i\phi_i} \, .
\end{equation}
The space of solutions is then dictated by which of the $\lbrace \eta_i, \alpha_i \rbrace$ is nonzero.

\subsection{Extremal solution}
Having shown that such exact scaling solutions only exist in theories with an exponential scalar potential and gauge coupling function, we rewrite the action and establish parametric definitions in terms of Lagrangian parameters. Our action is
\begin{equation}\label{action}
\begin{split} S = -\frac{1}{16\pi G_D} \int d^{D}x \sqrt{-g} \Big(R&+e^{\alpha\phi}F_{\mu\nu}F^{\mu\nu} +\frac{1}{2}(\partial\phi)^2 -V_0e^{\eta\phi}\Big) \, , \\ 
\end{split}
\end{equation}
Withou the flux term, this theory describes a consistent sphere truncation of a higher-dimensional supergravity to gravity coupled to a single scalar \cite{cvetic2}. Our ansatz, once more, is
\begin{equation}\label{scaling}
\begin{split}
ds^2&=-C_1r^{\beta}dt^2+\frac{dr^2}{C_1r^{\beta}}+C_3r^{\gamma}dx^i\cdot dx^i \\ 
\phi(r)&=C_2\ln{r}+\phi_0 \\ 
A_t'(r)&=\frac{\rho}{r^{\alpha C_2 + \gamma \frac{D-2}{2}}}\\
\end{split}
\end{equation}
As $\phi_0$ and $C_3$ can be eliminated by rescaling $r$ and $x^i$, respectively, we will set $\phi_0=0$ and $C_3=1$. Then the parameters of our ansatz, $\lbrace C_1, C_2, \rho, \beta, \gamma\rbrace$ are given in terms of the physical parameters of the theory, $\lbrace V_0, \eta,\alpha \rbrace$ as follows:
\begin{equation}\label{parameters}
\begin{split}
\beta &= 2-\frac{2(D-2)(\alpha+\eta)}{(\alpha+\eta)^2+2(D-2)}\eta \\
\gamma &= \frac{2(\alpha+\eta)^2}{(\alpha+\eta)^2+2(D-2)} \\
C_2 &= -\frac{(D-2)}{\alpha+\eta}\gamma \\
\rho^2 &= \frac{V_0}{2}\frac{2-\eta^2-\alpha\eta}{2+\alpha^2+\alpha\eta} \\
C_1 &= \frac{V_0 ((\alpha+\eta)^2+2(D-2))^2}{(D-2)(2+\alpha^2+\alpha\eta)(2(D-2)+\alpha^2(D-1)-\eta^2(D-3)+2\alpha\eta)}\\
\end{split}
\end{equation}
As we have no scale invariance, $C_1$ receives a constant rescaling as we rescale $r$. For order one radii, the validity of our classical analysis demands that $C_1$ is small. This is essentially the phenomenological version of taking the large N limit: for an action without fluxes descendant from string theory, there is no clear concept of what N is, and instead we just insist that the gravity theory is classical. On the field theory side, this guarantees that the density of degrees of freedom is large. 

Without loss of generality we restrict $\eta>0$, and equation (\ref{pot}) then implies that $V_0>0$ and $\phi$ must diverge to positive infinity at the horizon. 

We elucidate the content of these expressions by noting the following:
\begin{itemize}
\item When $\eta=0$, $\beta=2$: the scalar potential is constant, and our solutions become scale invariant. This framework enables us to consider AdS$_D$ ($\alpha \rightarrow +\infty, \gamma=2$, no flux, $\phi$ constant), AdS$_2\times\mathbb{R}^{D-2}$ ($\alpha \rightarrow 0, \gamma=0$, flux through $\mathbb{R}^{D-2}$, $\phi$ constant), and the modified Lifshitz solution ($\alpha$ arbitrary, $0 \leq \gamma \leq 2$, flux through $\mathbb{R}^{D-2}$, $\phi \sim \ln{r}$) as formal limits.

\item In order for $\phi(r) \rightarrow +\infty$ for small $r$ and the flux to be real, the bound on $\alpha$ in terms of some fixed $\eta$ is 
\begin{equation}\label{alphabound}
-\eta \, < \, \alpha \, < \, \frac{2}{\eta}-\eta 
\end{equation}
The lower bound says that $\beta \leq 2$, where saturation occurs only if the potential is constant ($\eta=0$). 

\end{itemize}

To be certain that this scaling solution is within the domain of validity of a classical gravitational treatment, we study the singularity structure of the spacetime. Calculation of the Ricci scalar, squared Ricci tensor and Kretschmann invariant reveal a curvature singularity at $r=0$ for the general scaling solution:
\begin{equation}
\begin{split}
R &= A_1 r^{\beta-2} \\
R_{\mu\nu}R^{\mu\nu} &= A_2 (r^{\beta-2})^2 \\
R_{\mu\nu\lambda\sigma}R^{\mu\nu\lambda\sigma} &= A_3(r^{\beta-2})^2 \, , \\
\end{split}
\end{equation}
where $A_i = f_i(\beta,\gamma)$ that can never simultaneously vanish. This reproduces the result that the extremal Lifshitz metric ($\beta=2$) has constant, finite curvature everywhere, but also tells us that spacetimes with $\beta < 2$ are smooth at large $r$, and in fact have asymptotically vanishing curvature invariants. We will return to this important fact in the next section, where we begin to discuss domain walls as holographic spacetimes. As for the singularity at $r=0$, we will show the near-extremal generalization of this solution presently, shielding the singularity in the usual manner.\footnote{We also note that our solution, and its near-extremal generalization, have a pp singularity at the horizon which indicates geodesic incompleteness due to diverging tidal forces as measured by a freely falling observer. The Lifshitz spacetime is known to suffer from such a feature at its horizon as well; as there, one can accept such singularities in hopes that they have some stringy resolution \cite{Horowitz}.}

\subsection{Near-extremal solution}
This theory also admits a finite temperature generalization of our scaling solution, whereby one adds an emblackening factor to the metric that protects the singularity at the origin: now, 
\begin{equation}\label{nearext}
U(r)=C_1r^{\beta}\Big(1-\Big(\frac{r_h}{r}\Big)^{\omega}\Big)\, ,
\end{equation}
where $\omega = \beta-1+\gamma\frac{D-2}{2}$, and all other fields and parametric definitions remain unchanged. Preservation of the correct metric signature is ensured for $\beta>1$.  

One finds the temperature of the geometry via the usual analytic continuation to Euclidean space, where demanding periodicity of the time coordinate so as to avoid conical singularity at the origin gives
\begin{equation}
T=\frac{1}{4\pi}C_1 \omega r_h^{\beta-1} \, .
\end{equation}

The entropy per unit volume of the planar horizon is
\begin{equation}
s\equiv \frac{S}{V_{\mathbb{R}^{D-2}}}=\frac{1}{4 G_D}r_h^{\gamma(\frac{D-2}{2})} \, ,
\end{equation}
yielding an entropy density-temperature scaling relation,
\begin{equation}\label{stpowerlaw}
s\sim T^{\chi} \, , \quad \chi = \frac{(D-2)(\alpha+\eta)^2}{2(D-2)+(\alpha+\eta)(\alpha-(2D-5)\eta)} \, .
\end{equation}
Other thermodynamic quantities follow from differentiation of the entropy density, e.g. the specific heat is positive and of the same power in $r_h$ as the entropy density.

By definition, $\chi > 0$ implies $\beta > 1$ and vice versa, ensuring that the $T=s=r_h=0$ extremal limit is obtained smoothly as we lower the temperature. One can arrange for an infinite range of $\chi$ by changing the physical parameters of the theory. 

Let us quickly note two facts about this scaling:
\begin{enumerate}
\item $\chi=D-2$, its free field value, when $\alpha=\frac{1}{\eta}-\eta$, a $D$-independent result.
\item For a given $\eta$, there is a one-parameter family of theories which have linear specific heat, $\chi=1$, for which
\begin{equation}
\alpha=\frac{1}{D-3}\left(\eta(5-2D)+\sqrt{\eta^2(D-2)^2+2(D-2)(D-3)}\right)
\end{equation}
This $\alpha$ is consistent with the bound (\ref{alphabound}). Generically, this value of $\alpha$ appears to have no relation to any AdS$_3$ geometry of the gravity dual.
\end{enumerate}


How do we use gauge/gravity duality in this situation? In the usual AdS/CFT correspondence, the behavior of the bulk fields maps onto properties of the dual field theory living on the boundary. The area of the black hole horizon would correspond to the low temperature entropy of the theory whose IR physics is captured by the bulk dynamics at small $r$. When $\beta=2$, AdS is the natural vacuum of the theory, and so such an interpretation is possible, presuming an interpolating solution exists that patches the IR geometry onto an asymptotic AdS in the UV. This was found numerically in \cite{kachru} for the extremal modified Lifshitz solution.\footnote{In Appendix A, we find it for the near-extremal Lifshitz solution.}


But in the context of a theory which does not admit an AdS vacuum, this framework obviously cannot apply. A global solution, if it exists, patches the IR dynamics onto some other UV spacetime: one must ask what this spacetime is, and whether a correspondence can be set up which permits us to make such thermodynamic identifications of geometric quantities. 

This latter question is answered affirmatively as we turn to the solution which is the analog of AdS in this theory, namely the domain wall spacetime.

\section{Domain walls}
\setcounter{equation}{0}

Consider the parametric relations (\ref{parameters}). Suppose our theory was such that $\alpha=\frac{2}{\eta}-\eta$, and hence the scaling solution would have zero flux: then the metric would have $\beta=\gamma\equiv \gamma '$. When $\gamma '=2$, this is of course AdS. When $\gamma '\neq 2$, this is none other than a domain wall solution, familiar from various contexts and the obvious backbone of the DW/QFT correspondence. For instance, the noncompact part of the extremal near-horizon D-brane geometry has this form, as do extremal solitonic solutions in dimensionally reduced gauged supergravities. We show below that, in fact,  such a domain wall solution exists for all allowed sets of $\lbrace\alpha,\eta\rbrace$.

Ignoring supersymmetries, this spacetime can have, at most, Poincar\'e symmetry. Whereas asymptotically AdS solutions are gravity duals to conformal field theories, asymptotically domain wall solutions are gravity duals to theories with only a relativistic symmetry. The decrease in elegance in establishing a correspondence in this case is countered by the increase in the number of real-world systems to which this treatment is applicable.

We first review the details of domain walls with an eye toward establishing the holographic correspondence. In the process, we show why domain walls with $\gamma ' < 1$ are unfit for this purpose. 
\subsection{Geometry}
The most general single-scalar domain wall solution can be written in the form
\begin{equation}\label{DW}
\begin{split}
ds^2&=(A r)^{\gamma '} \eta_{\mu\nu}dx^{\mu}dx^{\nu} + \frac{dr^2}{(A r)^{\gamma '}} \\
\phi(r) &= C_2 ' \ln {A r} \\
\end{split}
\end{equation}
In our model, the parametric definitions are
\begin{equation}\label{DWparams}
\begin{split}
\gamma ' &= \frac{4}{2+\eta^2 (D-2)} \\
C_2 ' &= -\eta \gamma ' \frac{D-2}{2}\\
A^2 &=  \frac{4V_0}{D-2}\frac{1}{\gamma '(D\gamma '-2)}\, .\\
\end{split}
\end{equation}

The forms of $\gamma '$ and $C_2 '$ can be obtained from the scaling solution parameters $\gamma$ and $C_2$, respectively, by substituting $\alpha=\frac{2}{\eta}-\eta$ (i.e. the $\rho=0$ condition) into the expressions (\ref{parameters}). The form of $A^2$ has a similar relation to its counterpart, $C_1$, which is only manifest in a gauge in which the constant part of the scalar is chosen to vanish: defining a new radial coordinate
\begin{equation}
R = Ar
\end{equation}
and shifting the boundary coordinates as 
\begin{equation}
\lbrace t, x^i \rbrace \, \rightarrow \, A\lbrace t, x^i \rbrace
\end{equation}
we have the solution 
\begin{equation}
\begin{split}
ds^2&=A^2 R^{\gamma '} \eta_{\mu\nu}dx^{\mu}dx^{\nu} + \frac{dR^2}{A^2 R^{\gamma '}} \\
\phi(r) &= C_2 ' \ln {R} \\
\end{split}
\end{equation}
Indeed, $A^2=C_1 (\alpha\rightarrow \frac{4}{\eta}-\eta)$.

Let us stress that the aforementioned substitution for $\alpha$ is only a parametric manipulation designed to derive the form of the vacuum solution; the solution exists for \textit{any} values of $\lbrace \alpha , \eta \rbrace$ consistent with other physical principles mentioned elsewhere. 

One can perform a perturbative analysis in the parameter $\eta$ by considering $\eta \ll 1$, which means a potential that is effectively constant over a large range of $\phi$. Because $\eta$ enters quadratically in the exponent $\gamma '$ but linearly in $C_2 '$, the solution (\ref{DW}), to first order in $\eta$, looks like AdS with a slowly rolling scalar field . The scalar Klein-Gordon equation expanded about the AdS background to first order in $\eta$, 
\begin{equation}
\Box \phi = -V_0 \eta + O(\eta^2) \, ,
\end{equation}
reveals logarithmic behavior for $\phi$ near the boundary, to leading order in $r$. 

While this domain wall solution makes clear its relation to our scaling solutions and to Dp-brane geometries, we wish to briefly make contact with other literature on solitonic supergravity domain walls by rewriting this metric in new, conformally flat coordinates: a harmonic function on the (one-dimensional) transverse space multiplying a Minkowski line element plus a ``radial" term, describing a solitonic $D-2$ brane in a D-dimensional space.

Labeling the new transverse coordinate $y$, we can write the metric as
\begin{equation}
ds^2 = H(y)^x (\eta_{\mu\nu}dx^{\mu}dx^{\nu}+dy^2) \, ,
\end{equation}
which defines the harmonic function and radial coordinate as
\begin{equation}
(A r)^{\gamma '} = H(y)^{x} \, , \quad \pm dy = \frac{dr}{(Ar)^{\gamma '}} \, .
\end{equation}
Assuming that $\gamma ' \neq 1$, integrating gives
\begin{equation}
H(y) = (1 \pm my) \, , \quad x=\frac{\gamma '}{1-\gamma '}
\end{equation}
where we define
\begin{equation}
m=A(1-\gamma ')
\end{equation}
and have chosen the integration constant such that the constant in $H(y)$ is equal to one. The full solution, with the scalar field, is thus
\begin{equation}
\begin{split}
ds^2 &= H(y)^{\frac{\gamma '}{1-\gamma '}} (\eta_{\mu\nu}dx^{\mu}dx^{\nu}+dy^2) \,  \\
e^{\phi} &= H(y)^{-\eta\frac{D-2}{4}\frac{\gamma '}{1-\gamma '}}\, \\
\end{split}
\end{equation}
The likeness of this solution to the noncompact part of Dp-brane metrics along with the running scalar is no coincidence, because various dimensional reductions of 10- and 11-dimensional maximal supergravities give rise to a panoply of such solitons in the corresponding lower-dimensional supergravities \cite{lpt}. The interpolating structure is evident: the metric asymptotes to flat space as $y \rightarrow 0$, and to near-horizon form with the asymptotically flat part decoupled as $y \rightarrow \pm \infty$. Note the $\pm$ sign in $H(y)$, which enables one to construct domain walls with $\mathbb{Z}_2$ symmetry. 

Further defining the constant
\begin{equation}
\Delta = \eta^2-2\frac{D-1}{D-2}
\end{equation}
enables us to write the full solution as 
\begin{equation}
\begin{split}
ds^2 &= H(y)^{\frac{4}{(D-2)(\Delta+2)}} (\eta_{\mu\nu}dx^{\mu}dx^{\nu}+dy^2) \,  \\
e^{\phi} &= H(y)^{-\frac{2\eta}{\Delta+2}}\,  \\
m^2 &= \frac{-V_0 (\Delta+2)^2}{2\Delta} \\
\end{split}
\end{equation}
This matches the solution in \cite{cvetic2}, for example. 

When $\gamma ' =1$, we instead have the relation 
\begin{equation}
H(y)^x = Ar = e^{\pm Ay} \, ,
\end{equation}
again with an integration constant chosen for simplification. Now,
\begin{equation}
\eta^2 = \frac{2}{D-2} \, , \quad A = \frac{2}{D-2}\sqrt{V_0} \, , \quad \Delta = -2
\end{equation}
and the full solution reads
\begin{equation}
\begin{split}
ds^2 &=e^{\pm \frac{2}{D-2}\sqrt{V_0} y} (\eta_{\mu\nu}dx^{\mu}dx^{\nu}+dy^2) \, , \\
\phi &= \mp \sqrt{\frac{2V_0}{D-2}} y  \\
\end{split}
\end{equation}

Returning to the form (\ref{DW}), one sees that, at least for $\gamma ' > 1$, the domain wall spacetime is well-suited for holographic correspondence. In accord with the discussion in section 2, it is well-behaved at large $r$, and more importantly, it has a boundary in the AdS sense. Specifically, the coordinate time to $r=\infty$ along a null (timelike) geodesic is finite (infinite), despite an infinite proper radial distance to the boundary from any point in the interior:
\begin{equation}
\begin{split}
\int^{t_f}_{t_i} dt = \int^{\infty}_{r_i} \frac{dr}{r^{\gamma '}} &= \text{finite} \\
\int ds = \int^{\infty}_{r_i}\frac{dr}{r^{\frac{\gamma '}{2}}} &= \infty \\
\end{split}
\end{equation}

Crucially, the domain wall spacetime is conformal to AdS: 
\begin{equation}\label{ads}
ds^2_{AdS} = r^{\gamma ' -2} ds^2_{DW} = A r^{2(\gamma '-1)} \eta_{\mu\nu}dx^{\mu}dx^{\nu} + \frac{dr^2}{A r^{2}}
\end{equation}
One should think of this metric as AdS in ``interpolating coordinates", where the value of $\gamma '$ determines whether the boundary lies at $r\rightarrow\infty$ or $r=0$. That is to say, $\gamma ' = 0$ gives AdS in conformally flat coordinates with a boundary at $r=0$, $\gamma ' = 2$ gives AdS in coordinates with the boundary at $r \rightarrow \infty$, and other values on either side of $\gamma ' = 1$ simply stretch one of these limiting spacetimes.\footnote{The value $\gamma ' = 1$ is once again special, as the metric is not AdS; we recognize this from the cases of the near-horizon metrics of NS5 and D5-branes of type II supergravity, a connection we discuss more later.}

These interpolating coordinates provide a convenient heuristic to understand why domain walls with $\gamma ' < 1$ do not have a boundary. Consider the construction of such domain walls by multiplying the AdS metric (\ref{ads}) by the conformal factor, $\Omega (r)=r^{2-\gamma '}$. This vanishes at $r=0$. Therefore, if we start from an AdS metric with the boundary at $r=0$, the map to the domain wall spacetime will not be faithful because the infinite volume near the boundary is cancelled by the degenerating conformal factor.

In addition to not having a boundary, the extremal $\gamma ' < 1$ domain walls have ill-defined thermodynamic properties. Generally, the local nature of near-horizon physics tells us that a physically acceptable extremal solution should be viewed as a zero temperature limit of a near-extremal solution, particularly if it is singular \cite{gubser2}. Presenting the finite temperature domain wall solution,
\begin{equation}\label{DWfiniteT}
ds^2 = -(Ar)^{\gamma '}(1-(\frac{r_h}{r})^{\omega '}) dt^2 + \frac{dr^2}{(A r)^{\gamma '}(1-(\frac{r_h}{r})^{\omega '})}+(Ar)^{\gamma '} dx^i \cdot dx^i \, ,  
\end{equation}
where
\begin{equation}
\omega ' = \frac{D}{2}\gamma '-1 \, ,
\end{equation}
we see that as $\gamma ' \rightarrow \frac{2}{D}$, $A$ diverges and the emblackening factor becomes identically one for all temperatures. When $\gamma ' < \frac{2}{D}$, $A$ becomes imaginary, so there is clearly some region of $\gamma ' < 1$ of which we should be suspicious.

We will shortly have more to say about why $\gamma ' < 1$ domain walls should be rejected as unphysical. 

\subsection{Holography}
Just as the AdS/CFT correspondence was fundamentally built on the physics of D3-branes and subsequently extended to apply to \textit{a priori} unrelated asymptotically AdS spaces, the validity of our duality for arbitrary Lagrangians of gauged supergravity-inspired form (\ref{action}), and arbitrary domain wall parameters (\ref{DWparams}), relies on the formal establishment of DW/QFT correspondence in connection to nonperturbative D-branes in string theory. We give a brief review; for details, see \cite{mald2,townsend,skenderis,bergshoeff}. 

The basic idea is that for D-branes with $p<6$, there is a formally well-defined limit in which gravity decouples at intermediate energies, and therefore one can describe the worldvolume theory using the strong field dynamics of the supergravity solution. This theory, the toroidal reduction of $D=10$, $\mathcal{N}=1$ super-Yang-Mills to $p+2$ dimensions, has a dimensionful coupling, $g_{YM}$, that runs with scale. This maps to a variable bulk dilaton. Consider the type II supergravity bosonic string frame action with the NS 2-form set to zero,
\begin{equation}
S=\frac{1}{(2\pi)^7l_s^8}\int d^{10}x\sqrt{-g}\left(e^{-2\phi}(R+4(\partial\phi)^2)-\frac{1}{2(p+2)!}F_{p+2}^2\right)
\end{equation}
A Dp-brane is electrically charged under the RR field strength as $F_{p+2}=dA_{p+1}$. The D-brane solution is
\begin{equation}
\begin{split}
ds^2&=\frac{1}{H_{p}(r)^{1/2}}ds^2_{M_{p,1}}+H_{p}(r)^{1/2}(dr^2+r^2d\Omega_{8-p}^2)\\
e^{\phi}&=g_sH_p(r)^{(3-p)/4}\\
A_{p+1}&=g_s^{-1}(H_{p}(r)^{-1}-1)\\
\end{split}
\end{equation}
with H a harmonic function on the transverse $(9-p)$-dimensional space, chosen as
\begin{equation}
H_p(r)=1+\frac{c_p g_s N l_s^{7-p}}{r^{7-p}}
\end{equation}
where $c_p$ is a $p$-dependent constant chosen to satisfy Maxwell's equation, $d\star F_{p+2}=0$. N is the quantized RR flux through $S^{8-p}$.

We now take the low energy, $l_s \rightarrow 0$ limit, keeping fixed both the energy of a stretched string ending on one of the stack of D-branes, as well as the coupling $g_{YM}^2 N$, where $g_{YM}^2=g_s l_s^{p-3}$. We also take $g_s \rightarrow 0$, hence working at tree level in the closed loop string perturbation expansion. Doing so reveals that, as for $p=3$, we decouple the asymptotically flat region and end up with a near-horizon geometry with metric 
\begin{equation}
ds^2=\left(\frac{r^{7-p}}{c_p g_s N l_s^{7-p}}\right)^{1/2}ds^2_{M_{p,1}}+\left(\frac{c_p g_s N l_s^{7-p}}{r^{7-p}}\right)^{1/2}(dr^2+r^2d\Omega_{8-p}^2)
\end{equation}
This geometry is a warped product of a $p+2$-dimensional domain wall with an $(8-p)$-sphere with an $r$-dependent radius. For $p<3$, the curvature is well-behaved at small $r$, but the effective string coupling $e^{\phi}$ blows up; for $p>3$, the situation is opposite.\footnote{The domain wall and sphere parts of the metric contribute to the curvature divergence in concert, so that one need only know where the sphere becomes small to know where the curvature blows up.}

The domain wall is conformal to AdS, and in fact, passing to the conformal frame -- known as the ``dual frame" in the DW/QFT literature -- gives us a product space AdS$_{p+2}\times S^{8-p}$:
\begin{equation}\label{conf}
\widetilde{g}_{\mu\nu}= (Ne^{\phi})^{\frac{2}{p-7}}g_{\mu\nu} \sim (g_{YM}^2 N)^{-1}r^{\frac{3-p}{2}}g_{\mu\nu} 
\end{equation}
and so, defining a new so-called ``horospherical" radial coordinate
\begin{equation}\label{uvar}
u^2 \sim  (g_{YM}^2 N)^{-1} r^{5-p}
\end{equation}
this metric reads as\footnote{The one exception to this is the case $p=5$, for which the variable $u$ is ill-defined and the conformal geometry is not AdS$_{7}\times S^{3}$, but rather $M_{5,1}\times \mathbb{R}\times S^3$.}
\begin{equation}
\widetilde{ds}^2 \sim u^2 ds^2_{M_{p,1}}+\frac{du^2}{u^2}+d\Omega_{8-p}^2
\end{equation}
This suggests identification of the radial coordinate $u$ with the energy scale of the worldvolume SYM theory, naturally incorporating the energy-distance relation of Dp-brane supergravity probes \cite{polchinskipeet}. This frame allows a simple sphere reduction ansatz, in which the flux is through the sphere and the D=$(p+2)$-dimensional action consists only of the universal sector with gravity and the scalar alone. Passage back to the $(p+2)$-dimensional Einstein frame reveals a domain wall metric.

One can calculate the near-extremal entropy of the $(p+2)$-dimensional domain wall solution compactified on a torus with $S^1$ radii $L$,
\begin{equation}
S = \frac{A}{4G_D}\sim L^p N^2  (g_{YM}^2N)^\frac{p-3}{5-p}u^\frac{9-p}{5-p} \, ,
\end{equation}
which is the correct result \cite{mald2}. Written in terms of an effective dimensionless Yang-Mills coupling 
\begin{equation}\label{geff}
g_{eff}(E)^2=g_{YM}^2NE^{p-3}\, ,
\end{equation}
and identifying $E\sim u$, the entropy is
\begin{equation}\label{braneentropy}
S\sim L^p N^2 E^p(g_{eff}(E))^{\frac{p-3}{5-p}} \, .
\end{equation}
In this form, the departure from conformality when $p\neq 3$ is clearest.

We have only sketched the derivation, having been cavalier about factors of $l_s$ and left out the conformal frame definitions of $\phi$ and $F_{p+2}$. What we wish to emphasize are the following three things: 
\begin{enumerate}
\item The scalar field is nonconstant in the dual frame, an obvious fact since the conformal transformation on the metric leaves the scalar alone.
\item  These brane solutions are simply special (and $\frac{1}{2}$-supersymmetric) cases of the domain walls we considered in section 3, where the value of $\eta$ is given in terms of $p=D-2$. 

Specifically, one can show that in the language established in section 3, the domain walls that come from D-branes have \cite{townsend} 
\begin{equation}\label{gammaprimep}
\gamma ' = \frac{2(9-p)}{p^2-7p+18} 
\end{equation}
Alternatively, they are the vacua of theories with 
\begin{equation}
\eta^2=\frac{2(p-3)^2}{p(9-p)}
\end{equation}
As a check, we note that for $p=3$, $\gamma ' =2$ and $\eta=0$ as required. We also point out that as anticipated earlier, the case $p=5$ gives a 7-dimensional domain wall with $\gamma ' =1$, which is not conformal to AdS$_7$. 

We present the fourth reason why domain walls with $\gamma ' < 1$ are not suitable for holography: when $p > 5$, for which $\gamma ' < 1$, there is not a well-defined brane decoupling limit.\footnote{The case of $\gamma ' =1$ itself -- that is, fivebranes -- does admit a decoupling limit, though it requires separate analysis. D0 branes also have $\gamma ' =1$, but our treatment of domain walls does not apply to two dimensions in which several of our formulae break down; from the standpoint of the D0-brane theory itself, the duality is not problematic.} All branes for which there is such a limit have $\gamma ' \geq 1$, by inspection of (\ref{gammaprimep}). Therefore, when $\gamma ' < 1$ one has no right to make an extension from the discretuum of brane solutions to the continuum of domain wall solutions and expect a holographic duality to hold.

\item One can apply this analysis to near-extremal D-branes too: start instead from the 10-dimensional finite temperature D-brane solution, 
\begin{equation}
ds^2=\frac{1}{H_{p}(r)^{1/2}}\left(-f(r)dt^2+dx^i\cdot dx^i\right)+H_{p}(r)^{1/2}\left(\frac{dr^2}{f(r)}+r^2d\Omega_{8-p}^2\right)
\end{equation}
where
\begin{equation}
f(r)=1-\left(\frac{r_h}{r}\right)^{7-p}
\end{equation}
Performing the above conformal transformation (\ref{conf}), and transforming once again to the $u$ coordinate (\ref{uvar}), yields 

\begin{equation}
\widetilde{ds}^2 \sim u^2 (-f(u)dt^2+dx^i\cdot dx^i)+\frac{du^2}{u^2f(u)}+d\Omega_{8-p}^2
\end{equation}
where
\begin{equation}
f(u)=1-\left(\frac{u_h}{u}\right)^{\frac{2(7-p)}{5-p}}
\end{equation}
The noncompact part of this geometry is at finite temperature, but is not AdS-Schwarschild. Upon reduction to D dimensions and passage back to the Einstein frame, one obtains the D-dimensional finite temperature domain wall solution (\ref{DWfiniteT}) with
\begin{equation}
\omega ' = \frac{2p(7-p)}{p^2-7p+18}
\end{equation}
This is positive for all $0<p<7$. As a check, we note that for $p=3$, $\omega=4$, corresponding to AdS$_5$-Schwarzchild.
\end{enumerate}

Having elaborated on the relevant properties of domain walls, we return to the goal at hand: construction of interpolating solutions between the finite temperature scaling solution and the domain wall. 

\section{Numerical construction of interpolating solution}
\setcounter{equation}{0}

The interpolation we construct connects the two spacetimes (\ref{scaling}) and (\ref{DW}), along with their associated scalar and gauge fields. This is done numerically, using a shooting technique. 

Let us summarize the method. The equations of motion fix the relation between fields at the horizon, $r=r_h$, save for some number of free parameters whose horizon values determine the full set of initial data. One must also use symmetries of the horizon metric to eliminate gauge degrees of freedom. Thusly, we can vary the temperature of the black hole and, ultimately, the boundary conditions at large $r$ by varying the horizon values of fields.

As a numerical matter, one must start integration at some small distance outside the horizon to seed the perturbations, so we develop the fields in a power series about the horizon. Integrating out to infinity, one finds that for sufficiently low temperatures, the metric looks like the finite temperature scaling solution over a large but finite range of $r$, after which the growing perturbations become large enough to backreact upon the metric and induce its asymptotic form. 

Consider the solution at the horizon. By definition of the horizon, and by separate rescalings of $t$ and the boundary coordinates $\lbrace x^i \rbrace$, we can fix
\begin{equation}
U(r_h)=0\, , \quad U'(r_h)=V(r_h)=1 \, .
\end{equation}
(This gauge is only appropriate for nonzero temperatures.) We expand the fields to second-order in the expansion parameter $\epsilon$,
\begin{equation}
\begin{split}
U(r_h+\epsilon)&\approx\epsilon+\frac{u_2}{2}\epsilon^2 +\ldots \\
V(r_h+\epsilon)&\approx1+v_1\epsilon+\frac{v_2}{2}\epsilon^2+\ldots\\
\phi(r_h+\epsilon)&\approx\phi_0+\phi_1\epsilon+\frac{\phi_2}{2}\epsilon^2+\ldots\\
\end{split}
\end{equation}
with Maxwell's equation determining the field strength to arbitrary order in terms of these expansions,
\begin{equation}
A_t'=\frac{Q}{e^{\alpha\phi}V^{\frac{D-2}{2}}}
\end{equation}
Then the field equations give the following relations in terms of two free parameters, $\lbrace \phi_0, Q \rbrace$:
\begin{equation}
\begin{split}
v_1 &= \Big(\frac{2}{D-2}\Big)\Big(V_0 e^{\eta\phi_0}-2Q^2e^{-\alpha\phi_0}\Big) \\
\phi_1 &=-\left(V_0\eta e^{\eta\phi_0} + 2Q^2\alpha e^{-\alpha\phi_0}\right) \\
v_2 &= \frac{1}{2}v_1^2-\Big(\frac{1}{D-2}\Big)\phi_1^2 \\
\phi_2 &=\Big(\frac{D-2}{2}\Big)\Big(2v_2-v_1^2\Big)\cdot\frac{v_1}{\phi_1} \\
u_2 &=\Big(-V_0\eta^2\phi_0e^{\eta\phi_0}+2Q^2\alpha e^{-\alpha\phi_0}(\alpha\phi_1 + (D-2)v_1)-\frac{D-2}{2}v_1-2\phi_2\Big)\cdot\frac{1}{\phi_1} \\
\end{split}
\end{equation}
At large $r$, the field equations near the domain wall boundary dictate the falloff of the fields as
\begin{equation}\label{dwfalloff}
\begin{split}
U&\approx U_c r^{\gamma '}+\ldots \\
V&\approx V_c r^{\gamma '}+\ldots\\
\phi&\approx C_2' \ln{r} + \phi_c+\ldots\\
A_t &\approx \mu + \frac{\rho}{r^{\alpha C_2' + \gamma '\frac{D-2}{2}}}+\ldots \\
\end{split}
\end{equation}
where we have kept only a handful of leading and subleading terms. We are free to normalize our horizon coordinates such that the asymptotic domain wall is Poincar\'e symmetric, $U_c=V_c$. (We do so with the understanding that the entropy density scales accordingly -- see Appendix A for a deeper discussion of this issue). It is our task, then, to show this falloff numerically.

The constants $\lbrace U_c, \phi_c \rbrace$ in the numerical large $r$ asymptotic region will not necessarily be equal to those in (\ref{DW}): we need to present a more general domain wall solution which allows us to account for differences in gauge between the horizon and infinity. Taking $r \rightarrow \frac{r}{c}$ and normalizing the boundary coordinates appropriately gives a solution 
\begin{equation}\label{DWnumerics}
\begin{split}
ds^2&=c^{2-\gamma '}(A r)^{\gamma '} \eta_{\mu\nu}dx^{\mu}dx^{\nu} + \frac{dr^2}{c^{2-\gamma '}(A r)^{\gamma '}} \\
\phi(r) &= C_2 ' \ln {Ar} - C_2 ' \ln {c} \\
\end{split}
\end{equation}
$c$ is an unconstrained gauge parameter which relates asymptotic constants in the simulation. Solving for c yields the relation
\begin{equation}\label{consts}
U_c = A^2 e^{-\left(\frac{2-\gamma '}{C_2 '}\right)\phi_c}
\end{equation}

If we set $\eta=0$, then we will interpolate between the modified Lifshitz solution and AdS. This was done in \cite{kachru} at zero temperature by solving for the exact form of the perturbation to linear order. In appendix A, we construct the analogous numerical solutions for the finite temperature modified Lifshitz solution, showing that at successively lower temperatures, the power law relation (\ref{stpowerlaw}) is increasingly obeyed. The lack of Poincar\'e invariance in the zero temperature solution implies the presence of an analytically unknown coefficient in the entropy due to the stretching of the spacetime between the horizon and infinity; but as the dimensionless temperature $\hat{T}\rightarrow 0$, this coefficient stabilizes to its zero temperature value, and power law behavior is observed. 

Here we present the results of one simulation with $\eta \neq 0$. Specifically, we choose the set of parameters
\begin{equation}
D=4\, , \quad \eta=\frac{1}{2}\ , \quad \alpha=1\, , \quad V_0 = 6\, , \quad r_h=1 \, .
\end{equation}
This translates to small $r$ scaling solution parameters
\begin{equation}
\beta=\frac{38}{25}\, , \quad \gamma=\frac{18}{25}\, \quad C_2=-\frac{24}{25}\, , \quad \rho=\sqrt{\frac{210}{196}}\, , \quad C_1 =\frac{1875}{434}
\end{equation}
and large $r$ domain wall parameters
\begin{equation}\label{DWpar}
\gamma ' = \frac{8}{5}\, , \quad C_2' = -\frac{4}{5}\ , \quad A^2=\frac{75}{44}\, .
\end{equation}

The interpolation is shown in figure 1. The upper plot shows the metric turnover: for a large but finite range in $r$, $V(r)$ scales as $r^\gamma$ as seen in the slope of the plot. (Though we did not show it, one can of course directly extract the IR power law relation for $U(r)$ as well.) At large $r$, both metric components scale with domain wall exponent $\gamma ' = \frac{8}{5}$, where the turnover on the $V(r)$ plot happens around $\ln{r} \sim 5$.

The lower plots show the scalar and field strength, in which we plot elucidating combinations of the fields and $r$ based on the asymptotic scalings described above. In particular, in the large $r$ domain wall region, we have
\begin{equation}
A_t'(r) \sim r^{-\frac{4}{5}} \sim e^{\phi}
\end{equation}
by plugging the values (\ref{DWpar}) into the falloff (\ref{dwfalloff}). 

One can also extract the value of $\phi_c$ from the scalar plot and compare it to the value $U_c$ (not shown); the numerical values for this trial, to six significant figures, are
\begin{equation}
U_c = 19.4192 \, , \quad \phi_c = 4.86591 \, ;
\end{equation}
the relation (\ref{consts}) is satisfied to great accuracy. 

\begin{figure}[hb!]
\begin{centering}
  \includegraphics[scale=.6]{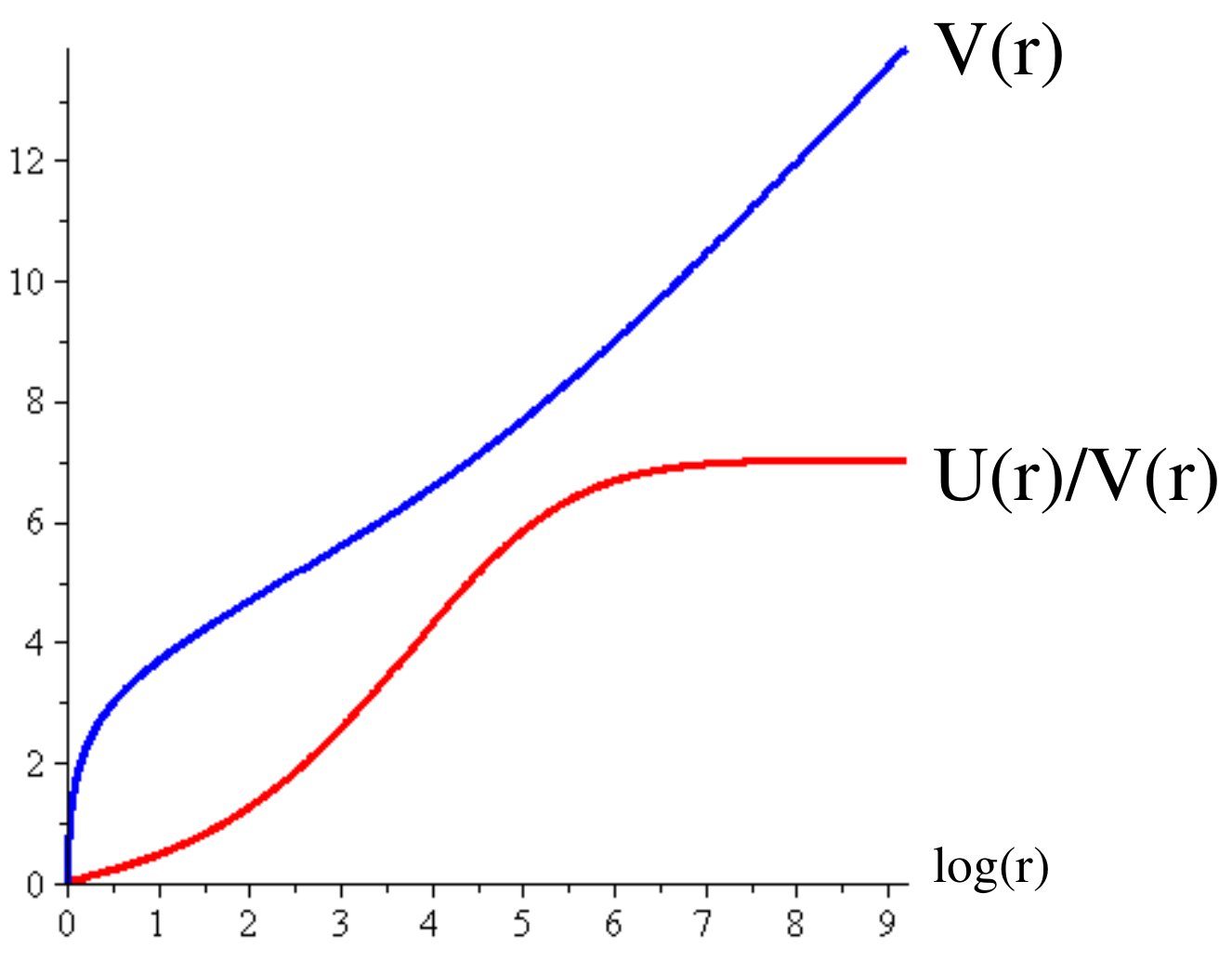}\\
\end{centering}
\begin{centering}
  \includegraphics[scale=.6]{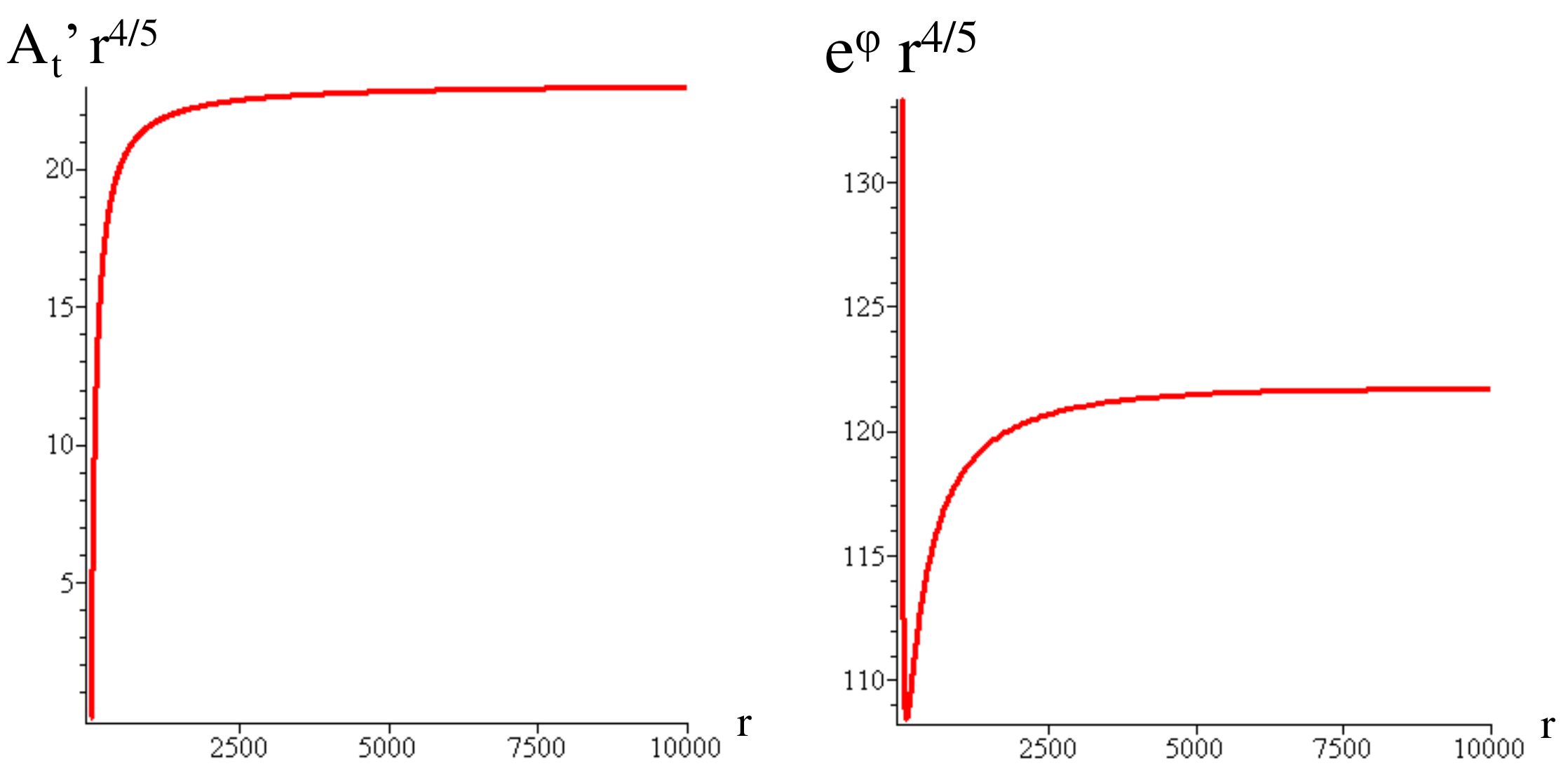}\\
\end{centering}
\caption{Behavior of fields in the interpolating solution. a. \textit{Upper}: The metric component $V(r)=g_{ii}$, as well as the ratio $\frac{U(r)}{V(r)}=\frac{-g_{tt}}{g_{ii}}$, rescaled to fit on the same graph. The turnover in the slope of the former curve, and the flatness of the latter, indicate the domain wall asymptotics. b. \textit{Lower left}: The field strength, multiplied by its asymptotic domain wall power of $r$. c. \textit{Lower right}: The exponentiated scalar field, multiplied by its asymptotic domain wall power of $r$.}
\end{figure}

\newpage

\section{Generalized scale invariance}
\setcounter{equation}{0}

Having shown that the scaling solutions (\ref{scaling}) can indeed asymptote to domain walls (\ref{DW}) to form a global geometry, we ask what domain wall holography can tell us about this solution.

Our scaling solution is not scale-invariant for $\eta \neq 0$, but its thermodynamics are governed by power law relations. One can ask what aspects of this IR scaling behavior can be explained in connection to the symmetries of the domain wall. More specifically, the UV domain wall spacetime and hence the theory at hand has Poincar\'e symmetry, but also the generalized conformal structure explained earlier. It is worthwhile to consider if the IR scaling solution has any such ``hidden" symmetry as well: although the scaling solution is not scale invariant, one might expect it to possess a ``generalized scale invariance," manifest in the conformal frame defined by the domain wall.

To motivate this prospect, one may think of the flow from the domain wall to the scaling solution as a symmetry-breaking flow: the IR flux breaks the Lorentz symmetry of the asymptotic domain wall. But in Einstein frame, one cannot easily tell whether this particular solution leaves the domain wall's generalized scale invariance intact in the IR. That is to say, the presence of the IR flux might be expected to preserve that part of the generalized conformal structure which is independent of the Lorentz symmetry-breaking. 

We now show that this is indeed the case, by passage to the conformal frame. 

The DW/QFT correspondence says that we should use the UV domain wall geometry to determine the conformal factor, as discussed in section 3. This is easily done, writing the factor in terms of $r$ and converting it to an expression in terms of the field $\phi$. With the conformal factor in hand, we note that $\phi$ has a different $r$-dependence in the IR scaling solution, and therefore, the IR metric receives a different multiplicative factor in terms of $r$. Altogether, we have the new global conformal frame metric which interpolates between two scale-invariant metrics: a modified Lifshitz solution in the IR and an AdS plus rolling scalar solution in the UV.

For clarity, we reproduce the interpolating solution in Einstein frame as follows. The metric is
\begin{equation}\label{interp}
\begin{split}
ds^2_{IR}&= -C_1 r^{\beta}f(r)dt^2+\frac{dr^2}{C_1 r^{\beta}f(r)}+r^{\gamma}dx^i\cdot dx^i \\
ds^2_{UV}&= c^{2-\gamma '}(Ar)^{\gamma '}\eta_{\mu\nu}dx^{\mu}dx^{\nu}+\frac{dr^2}{c^{2-\gamma '}(Ar)^{\gamma '}}\\
\end{split}
\end{equation}
and $f(r)$ is defined in (\ref{nearext}).
The scalar field is
\begin{equation}
\phi(r) =\begin{cases} & C_2\ln{r}\, \phantom{A} \quad  r\rightarrow 0 \\ & C_2'\ln{\frac{A}{c}r}\,  \quad r\rightarrow \infty \\ \end{cases}
\end{equation}
The field strength is
\begin{equation}
A_t'(r)=\frac{\rho}{e^{\alpha\phi}V^{\frac{D-2}{2}}}\sim\begin{cases} & \rho r^{-(\alpha C_2 + \gamma\frac{D-2}{2})}\phantom{'}\, \quad  r\rightarrow 0 \\ & \rho r^{-(\alpha C_2' + \gamma '\frac{D-2}{2})}\, \quad r\rightarrow \infty\\  \end{cases}
\end{equation}
We set $c=1$ for simplicity, remembering that a generic choice of horizon coordinate normalizations will change this value.

From $ds^2_{UV}$, define
\begin{equation}
\widetilde{ds}^2_{UV} = \Omega (r)_{UV}ds^2_{UV}
\end{equation}
where
\begin{equation}
\Omega (r)_{UV}= \left(Ar\right)^{\gamma '-2}
\end{equation}
This will give us a UV AdS spacetime. In terms of $\phi$, this is proportional to the potential,
\begin{equation}
\Omega (\phi)= e^{\eta\phi}
\end{equation}
On the global solution, therefore, $\Omega (r)$ takes the following limiting forms:
\begin{equation}
\Omega(r)=\begin{cases} &r^{\beta-2}\phantom{(Ar)} \quad r\rightarrow 0 \\ & \left(Ar\right)^{\gamma '-2} \quad r\rightarrow \infty \\ \end{cases}
\end{equation}
where we have made use of algebraic relations $\eta C_2=\beta - 2$ and $\eta C_2' = \gamma ' -2$, by inspection of (\ref{parameters}) and (\ref{DWparams}), respectively.
Hence, the conformal frame metric is
\begin{equation}
\begin{split}
\widetilde{ds}^2_{IR}&= -C_1r^{2(\beta-1)}f(r)dt^2+\frac{dr^2}{C_1 r^2 f(r)}+r^{\beta+\gamma-2}dx^i\cdot dx^i\\
\widetilde{ds}^2_{UV}&=\left(Ar\right)^{2(\gamma '-1)}\eta_{\mu\nu}dx^{\mu}dx^{\nu}+\frac{dr^2}{(Ar)^2} \\
\end{split}
\end{equation}

Both the IR and UV metrics now have scale invariance, anisotropic in the former and isotropic in the latter. In other words, the general scaling solutions have a generalized scale invariance, manifest only in this conformal frame. Actually, the IR metric has its scale invariance broken by nonzero temperature, but the emergence of finite temperature Lifshitz behavior is still non-generic.

Because these solutions lie in the IR of a global solution asymptoting to a domain wall, and because one uses the domain wall alone to determine the Weyl factor $\Omega(\phi)$, this generalized scale invariance follows from the generalized conformal structure of the UV domain wall, itself grounded in the identical structure of nonconformal D-branes. 

To clarify the solution, we can define separate new radial coordinates in the IR and the UV which will restore the gauge choice $g_{tt}=-g^{rr}\neq g_{ii}$, thereby reinstituting horospherical coordinates in the UV AdS spacetime. (Note that as it stands, the UV metric is AdS in previously defined interpolating coordinates, (\ref{ads}).) Define the IR coordinate $u$ as

\begin{equation}
(\beta-1)u=r^{\beta-1}
\end{equation}
and the UV coordinate $R$ as
\begin{equation}
A(\gamma '-1)R=\left(Ar\right)^{\gamma '-1}
\end{equation}
Then the metric is
\begin{equation}\label{confmetric}
\begin{split}
\widetilde{ds}^2_{IR}&= -\widetilde{C}_{1}u^2f(u)dt^2+ \frac{du^2}{\widetilde{C}_{1}u^2f(u)}+\widetilde{C}_{3}u^{\widetilde{\gamma}}dx^i\cdot dx^i \\
\widetilde{ds}^2_{UV}&=-(\widetilde{A}R)^2dt^2+\frac{dR^2}{(\widetilde{A}R)^2}+(\widetilde{A}R)^2dx^i\cdot dx^i \\
\end{split}
\end{equation}
with conformal frame parameters
\begin{equation}\label{confparams}
\widetilde{\gamma}=\frac{\beta+\gamma-2}{\beta-1} \, , \quad \widetilde{C}_{1}=C_1(\beta-1)^2 \, , \quad \widetilde{C}_{3}=(\beta-1)^{\widetilde{\gamma}} \, , \quad \widetilde{A}=A(\gamma ' -1)
\end{equation}
and an emblackening function
\begin{equation}\label{emblack}
f(u)=1-\left(\frac{u_h}{u}\right)^{\widetilde{\omega}} \, , \quad \widetilde{\omega}=\frac{\omega}{\beta-1} 
\end{equation}
We note that the definitions (\ref{confparams}) guarantee small conformal frame curvatures, given small Einstein frame curvatures. 

The full solution includes the scalar field,
\begin{equation}\label{confscalar}
\phi =\begin{cases} &\widetilde{C}_{2}\ln{(u(\beta-1))} \quad r\rightarrow 0 \\ &\widetilde{C}_{2}'\ln{\widetilde{A}R}\phantom{\beta-1)} \quad r\rightarrow\infty \\ \end{cases}
\end{equation}
where 
\begin{equation}
\widetilde{C}_2=\frac{C_2}{\beta-1} \, , \quad \widetilde{C}_{2}'=\frac{C_2'}{\gamma '-1}
\end{equation}

To find the field strength in the new coordinates, we need to transform the two-form $F=\frac{1}{2}F_{\mu\nu}dx^{\mu}\wedge dx^{\nu}$, not just the tensor components. Doing so reveals an IR two-form
\begin{equation}\label{confgaugeir}
F_{IR}\sim u^{\frac{D-2}{2} \frac{\gamma}{\beta-1}}du\wedge dt
\end{equation}
and UV two-form
\begin{equation}\label{confgaugeuv}
F_{UV}\sim R^{2-D-\widetilde{C}_{2}'(\alpha-\frac{\eta}{2}(D-4))}dR\wedge dt
\end{equation}

The full conformal frame solution, then, is (\ref{confmetric}),  (\ref{confscalar}), (\ref{confgaugeir}), and (\ref{confgaugeuv}), with the associated parametric definitions. Note that when $\beta=\gamma '=2$ -- which would be the case for the theory with constant potential, giving rise to an interpolation between the modified Lifshitz solution and AdS -- the conformal frame and the Einstein frame are identical, and all of these formulae reduce to tautology.

The two radial coordinates $u$ (in the IR) and $R$ (in the UV) dualize to the energy scale of the field theory, each in its respective regime. If we are to consistently identify the bulk radial direction with the energy scale, then it must be true that the two coordinates smoothly interface at intermediate energies: when $u$ decreases, so must $R$, and both must have a strictly positive range. The coordinates share the relation
\begin{equation}\label{uRrelation}
u = \mathcal{A} R^{(\beta-1)/(\gamma ' -1)} \, ,
\end{equation}
where $\mathcal{A}=\frac{1}{\beta-1}(A^{2-\gamma '}(\gamma '-1))^{\frac{\beta-1}{\gamma ' -1}}$. When $\beta>1$ and $\gamma '>1$, the relationship is as required. 

By definition, the conformal frame solutions are those of the conformal frame action, with Lagrangian density
\begin{equation*}
\tilde{\mathcal{L}} = \sqrt{-\tilde{g}}e^{-\eta(\frac{D-2}{2})\phi}\Big(\tilde{R}+\Big(1+\frac{\eta^2}{2}(D-1)(D-2)\Big)\frac{\partial_{\mu}\phi\tilde{\partial}^{\mu}\phi}{2} + e^{(\alpha+\eta)\phi}F_{\mu\nu}\tilde{F}^{\mu\nu}-V_0\Big)
\end{equation*}
The tilde's in this action, as usual, represent quantities calculated using the conformal frame metric, $\tilde{g}_{\mu\nu}$. One can easily confirm that the gauge field scalings (\ref{confgaugeir}) and (\ref{confgaugeuv}), for example, satisfy Maxwell's equation,
\begin{equation}
\partial_{\mu}\left(\sqrt{-\tilde{g}}e^{(\alpha-\frac{\eta}{2}(D-4))\phi}F^{\mu\nu}\right)=0
\end{equation}

To summarize, the fact that we recover the finite temperature modified Lifshitz solution in the IR conformal metric (\ref{confmetric}) indicates that the dynamics of the dual relativistic field theory at low energies are determined in some fashion by those of scale-invariant fixed points.\footnote{We note, without clear understanding of its meaning, that for $\beta > 1$ and $\gamma ' > 1$ as here, we have $\tilde{\gamma}=\frac{2}{\tilde{z}}<2$, which is the $\tilde{z}>1$ condition of an Einstein frame Lifshitz gravity dual.}

Because this connection is a direct result of the inherent structure of the domain wall metric, it would be interesting to answer the question, ``what kinds of matter Lagrangians coupled to the universal Einstein-scalar sector can support solutions \textit{without} generalized scale invariance?" We elaborate on this in the conclusion.

\section{Domain wall holography and effective field theory}
\setcounter{equation}{0}
The solution we have constructed interpolates between two exact solutions of our single-exponential, domain wall gravity theory. We found ourselves restricted to such a theory upon demanding that the scaling ansatz exactly solve the theory defined by the general action,
\begin{equation}
\begin{split} S = -\frac{1}{16\pi G_D} \int d^{D}x \sqrt{-g} \Big(R&+f(\phi)F_{\mu\nu}F^{\mu\nu} +\frac{1}{2}(\partial\phi)^2 + \mathcal{V}(\phi)\Big) \, , \\ 
\end{split}
\end{equation}
As noted in the introduction, this is because the scaling ansatz, which models generic near-horizon behavior of zero entropy extremal solutions, singles out the term in some general $\mathcal{V}(\phi)$ that dominates at small $r$. Including subleading terms in $r$ in our ansatz -- essentially near-horizon corrections -- would turn on other contributions to the potential. 

For these reasons, our numerical work has broader suggestive power, specifically to theories with scalar potentials that are dominated by a single exponential term over a large range of field space but have an AdS critical point. We give a schematic example in the introduction, equation (\ref{poten}); more generally, we refer to the behavior
\begin{equation}\label{dwads}
\mathcal{V}(\phi) \sim \begin{cases} & -V_0e^{\eta\phi} \quad r\rightarrow 0 \\ & -V_0'\phantom{e^{\eta\phi}} \quad r\rightarrow \infty \\ \end{cases}
\end{equation}

Suppose that our theory was UV-completed with a full string/M-theory-derived potential with this behavior; the scaling solution of this paper would then be only approximate. If the scaling solution at small $r$ could be patched onto an approximate domain wall solution at intermediate $r$ whose only nonzero matter field is the scalar, then the interpolation between the domain wall and the AdS critical point at large $r$ would be guaranteed: the scalar would just roll toward its AdS critical point, in the manner of an RG flow. All other fields are turned off.  

Whether the interpolation could actually be constructed in this imaginary theory is a matter of numerics, not argument: one cannot assume the existence of an interpolating solution in a given theory without actually constructing it. But it seems reasonable that given a stringy effective action, a more complicated extremal solution to that theory which approaches our scaling solution in the near-horizon limit would interpolate to the approximate domain wall solution of that theory at intermediate energies. This should also be true for an ad hoc action with a scalar potential as in (\ref{dwads}). 

In fact, this sort of analysis leads one to consider DW/QFT as an effective holographic tool, applicable in settings far more general than domain wall supergravities, where we use ``effective" in the field theory sense. Our assertion is that even when the UV completion of some bulk theory is unknown, if that theory admits an approximate domain wall solution at some intermediate value of $r$ then one can use DW/QFT to develop a holographic map. 

Let us explore this idea.

The central tenet of effective field theory, informed by the philosophy of the renormalization group, is that physics at low energies should not be sensitive to physics at high energies. Naively, AdS/CFT seems to violate this idea: given an effective field theory, one needs to know something about its UV physics, namely whether the theory is conformal, in order to know whether it has a well-defined IR gravity dual. Of course this is an incorrect mode of thought, because gauge/gravity duality, it can be said, is not a right, but a privilege of conformal field theories (and their deformations and subsequent holographic extensions). There is no \textit{a priori} reason why every field theory should have a gravity dual. From what we know so far, one could argue that classifying the types of field theories which \textit{might} have gravity duals boils down to finding bulk spacetimes with boundaries, in which case the attendant isometries dictate the field theory symmetries.  

Actually, we can consider an elementary case in which the bulk has no AdS solution in the UV and we know how to treat it holographically: the case of a positive mass scalar field. As a bulk field dual to an irrelevant CFT operator, the scalar blows up at large $r$. At the radius at which its backreaction ruins the AdS asymptotics, we work with a radial cutoff, dual to working with an effective field theory below the corresponding energy scale. To incorporate the backreaction of the scalar would be to find the full UV completion of the theory; in its absence, one works with the effective AdS boundary and proceeds essentially as usual. 

Analogously, one is entitled to use DW/QFT in settings beyond those in which the domain wall geometry is a true vacuum of the theory -- one can work with a cutoff boundary at finite $r$, at which the bulk fields will have falloffs characteristic of a domain wall boundary. This is exactly what one does in the case of nonconformal D-branes, fundamental strings, and NS5 branes: the supergravity approximation cannot be trusted for large (or small) $r$, i.e. high (or low) energies. Of course, for some of these branes, the UV completion is given in terms of M-theory solutions in one greater spacetime dimension, which is interesting in and of itself: even in cases like the D4-brane where we know that the solution uplifts to an AdS$_7\times S^4$ vacuum in the UV, we can still define an effective ten-dimensional nonconformal holography. This M-theory resolution of strong effective ten-dimensional string coupling is obviously a deep and special case, but it alludes to the general possibility that even when a potential (borne from string/M-theory compactification or otherwise) has no AdS vacuum, DW/QFT can be used at intermediate energies. 

To rephrase, accepting the validity of nonconformal holography demands that it should have the same role as AdS holography in situations where the bulk must have a UV radial cutoff. In the absence of a continuum limit, even in AdS/CFT, one cannot be sure that holography is describing a theory that is not sick; this is no different in the nonconformal case and permits us to extend its regime likewise.

We can summarize when it is safe to use domain wall holography in theories that admit (at least approximate) domain wall vacua. 
\begin{enumerate}
\item When a phenomenological theory admits an exact domain wall solution, like our theory (\ref{action}), domain wall holography is on firm footing. Without a string/M-theory embedding, the theory is not quantumly well-defined; when an exact domain wall solution \textit{does} arise in a consistent truncation of string/M-theory, domain wall holography is valid even on the level of quantum corrections. 

In fact, our action (\ref{action}) can indeed be obtained in such a manner, where $\alpha$ and $\eta$ are fixed by the compactification. Specifically, \cite{cveticmain} showed that starting with the canonical $S^4, S^5$ and $S^7$ sphere compactifications of string and M-theory, one can deform these spheres by taking certain moduli to a limit in which the compactification on $S^n$ becomes one on $S^a \times \mathbb{R}^b$, where $a+b=n$. The resulting D-dimensional effective Lagrangian, suitably (and consistently) truncated to a single scalar, reads
\begin{equation}\label{stringy}
\begin{split} S = -\frac{1}{16\pi G_D} \int d^{D}x \sqrt{-g} \Big(R&+e^{-\sqrt{\frac{2}{D-2}}\phi}F_{\mu\nu}F^{\mu\nu} +\frac{1}{2}(\partial\phi)^2 -V_0e^{\sqrt{\frac{2}{D-2}}\phi}\Big) \, ; \\ 
\end{split}
\end{equation}
that is, $-\alpha=\eta=\sqrt{\frac{2}{D-2}}$. Unfortunately, our scaling solution does not solve these actions -- it reduces to the AdS$_2\times\mathbb{R}^{D-2}$ solution -- and so we make no use of them in this instance.

\item When a bulk action has a potential $\mathcal{V}(\phi)$ which has a large $r$ AdS vacuum, we can still use domain wall holography at low and intermediate energies if $\mathcal{V}(\phi)$ admits an approximate domain wall solution there, as exemplified in (\ref{poten}) and (\ref{dwads}). (Again, issues of UV completion come to bear on whether this is a microscopically allowed map.) The bulk fields that run with scale in the domain wall region eventually stabilize in AdS. For instance, in nonconformal $(p+1)$-dimensional super-Yang-Mills, the two-point function of a $\Delta=p+1$ operator dual to a massless scalar field, as shown in \cite{skenderis}, has scale-dependence determined by generalized conformal invariance as
\begin{equation}
\langle\mathcal{O}(x)\mathcal{O}(0)\rangle\sim N^2 (g_{eff}^2(|x|))^{\frac{p-3}{5-p}} \frac{1}{|x|^{2\Delta}}
\end{equation}
where $g_{eff}^2(|x|)=g_{YM}^2N|x|^{3-p}$, essentially as defined in (\ref{geff}). In the domain wall region, this function runs with scale; at the eventual AdS critical point, $g_{eff}$ runs smoothly to a constant and the conformal structure of the two-point function is recovered.
\end{enumerate}

In connection to the scaling solution studied in this paper, these arguments serve to suggest that we do not need to know whether the potential $\mathcal{V}(\phi)$ has an AdS vacuum at large $r$ in order to know something about the universal behavior of the dual field theory.

As a final point, the relation between higher-dimensional AdS solutions of M-theory and lower-dimensional domain wall solutions of string theory may be a more general aspect of the existence of DW/QFT. In \cite{skenderis2}, the authors show that any D-dimensional domain wall gravity theory with action
\begin{equation}
\begin{split} S = -\frac{1}{16\pi G_D} \int d^{D}x \sqrt{-g} \Big(R&+\frac{1}{2}(\partial\phi)^2 -V_0e^{\eta\phi}\Big) \, , \\ 
\end{split}
\end{equation}
can be derived by toroidal dimensional reduction of a pure AdS gravity, where the toroidal dimension is related to the value of the potential parameter $\eta$. This leads one to speculate that all holography is intimately connected to the existence of \textit{some} AdS vacuum, be it in the same spacetime dimension as the theory one is considering or otherwise. This would be an interesting conclusion that would generalize the way the strong coupling singularities of type IIA solutions are resolved.

\section{Discussion and prospects}
\setcounter{equation}{0}

To recapitulate, the scaling behavior of low temperature, relativistic quantum field theories with finite charge density with IR gravity duals (\ref{interp}) can be understood via domain wall holography and an inherited generalized scale invariance (broken by the finite temperature). The power law s-T relation of the bulk ansatz encodes the physics of a system with a unique zero temperature ground state, though the thermodynamic description breaks down due to a physical singularity in the extremal limit. 

We have numerically constructed the interpolating solution between the near-extremal scaling solution in the IR and the asymptotic domain wall vacuum in the UV, thus permitting the mapping of the near-horizon physics to the low-energy dynamics of the dual field theory. The formalism developed in \cite{skenderis,wisemanwithers} ensures that the holographic relation is faithful and thermodynamically well-defined. 

We also made some comments on the nature of domain wall holography for bottom-up actions with no connection to type II supergravity, delineating which domain walls are amenable to holography and arguing for an effective role of domain wall holography in settings where an exact domain wall vacuum does not exist. 


Let us emphasize that the generalized scale invariance of the IR solution followed, via the construction of the full interpolating solution, directly from the generalized conformal invariance of domain walls which is itself descendant of M-theory. The D4 brane of 10-dimensional, type IIA supergravity descends from the 11-dimensional solitonic M5-brane wrapped on the M-theory circle, for example, and similarly for the IIA fundamental string from the M2-brane. By using S- and T-duality on, say, the D4 brane supergravity solution, one can generate all branes of type II supergravity, including the D-branes of course; therefore, insofar as one defines their generalized conformal structure as the presence of a metric conformal to AdS$_{p+2}\times S^{8-p}$, the entire domain wall holography has a non-perturbative connection to M-theory in this way. Domain walls with no necessary relation to D-brane near-horizons should still be considered as a rung on this non-perturbative ladder, just as any asymptotically AdS spacetime can be treated holographically in its own right. 

It would be interesting to investigate what matter Lagrangians, when coupled to an Einstein-scalar sector with a domain wall vacuum, would break this generalized scale invariance. One might phrase this as follows. Suppose our potential is still given as
\begin{equation}
\mathcal{V}(\phi)=-V_0 e^{\eta\phi} \, .
\end{equation}
We know that the domain wall metric will be conformal to AdS via the conformal factor $\Omega(\phi)=e^{\eta\phi}$, so the Einstein frame metric 
\begin{equation}
ds^2 = -r^2e^{\eta\phi}dt^2 + \frac{dr^2}{r^2e^{\eta\phi}} + V(r)dx^i\cdot dx^i
\end{equation}
will have conformal frame scale invariance. What type of matter Lagrangian can support this metric?

It would also be worthwhile to find a string/M-theory embedding of our scaling solution, or a generalization of it. The string theory-derived effective action (\ref{stringy}) which we presented earlier could not accomodate it, but there are large families of similar actions of domain wall gravities, as laid out in \cite{cvetic3}. The domain wall would generally be supported by some number of scalars, which would be no impediment to use of  DW/QFT.

Lastly, it would clearly be nice to delve deeper into the thermodynamics of this system, though that has largely been done in the recent papers cited earlier \cite{charmousis, Lee:2010qs}. Perhaps adding fermionic degrees of freedom to the theory would be worthwhile as well.

\bigskip \bigskip 

\noindent
{\Large \bf Acknowledgments}\medskip

We thank Per Kraus for frequent guidance, crucial insight and consistent support. This work was supported in part by a Leo P. Delsasso Fellowship from the UCLA Department of Physics and Astronomy.

\bigskip \bigskip 

\noindent
\appendix
\section{Numerical study of finite temperature modified Lifshitz solution}
\setcounter{equation}{0}

\medskip

We clarify a lingering issue in the extraction of field theory thermodynamics from interpolating gravity solutions, as we present the numerical construction of the interpolating solution for a finite temperature black hole with IR Lifshitz scaling.\footnote{This work was done with Per Kraus.}

The motivation for doing so is that without patching the IR spacetime onto its UV AdS vacuum, the entropy density for the dual field theory can only be known up to an overall coefficient. The IR black hole provides the scaling behavior of the entropy, but because the spacetime is warped as the IR solution is patched onto an asymptotic AdS in the UV, the normalization of the horizon area generally receives a rescaling upon insisting that the AdS metric take its canonical, Poincar\'e-invariant form. In other words, if one defines the normalization of the boundary coordinates by using the AdS asymptotics, one must rescale the coordinates globally, and hence the area of the horizon will be rescaled by some constant. 

Consider the zero temperature interpolating metric, which we write as

\begin{equation}
ds^2=-U_0(r)dt^2+\frac{dr^2}{U_0(r)}+V_0(r)dx^i\cdot dx^i
\end{equation}

For an extremal black hole with Lifshitz scaling,
\begin{equation}
r \, \rightarrow\,  \frac{r}{\lambda} \, , \quad t \, \rightarrow \lambda t \, , \quad x^i  \, \rightarrow \lambda^{\frac{1}{z}} x^i 
\end{equation}
the IR behavior of the metric is determined by this scale invariance up to a constant:
\begin{equation}
ds^2=-\frac{r^{2}}{l^2}dt^2+\frac{l^2}{r^{2}}dr^2+\xi \Big(\frac{r}{l}\Big)^{2/z}dx^i\cdot dx^i \, ,
\end{equation}
where $\xi$ is a constant. The solution is not Poincar\'e invariant. As noted earlier, this interpolating solution was numerically constructed in \cite{kachru} for the $D=4$ extremal solution, where the authors solved the linearized perturbation equations. The metric goes as
\begin{equation}
U_0(r) = \left\{   \begin{array}{ccc} \left(r/l\right)^2  && r\rightarrow 0 \\
\left(r/L\right)^2   && r\rightarrow \infty \end{array}\right.
\end{equation} 
and
\begin{equation}
V_0(r) = \left\{   \begin{array}{ccc} \xi \left(r/l\right)^{2/z}  && r\rightarrow 0 \\
\left(r/L\right)^2   && r\rightarrow \infty \end{array}\right.
\end{equation} 
where $l$ and $L$ are the characteristic length scales of the Lifshitz and AdS geometries, respectively. $\xi$ is not fixed by any symmetry.

In thinking about the finite temperature solutions, one can view them as being ``glued in" to the zero temperature background, which is to say that the near-extremal black holes have extremal asymptotics:
\begin{equation}
ds^2=-\frac{r^{2}}{l^2}f(r)dt^2+\frac{l^2}{r^{2}}\frac{dr^2}{f(r)}+\xi \left(\frac{r}{l}\right)^{2/z}dx^i\cdot dx^i \, ,
\end{equation}
where $f(r)=1-\Big(\frac{r_h}{r}\Big)^{\omega}$ and $\omega=1+\frac{D-2}{z}$. In the context of an interpolating solution, this is only strictly true in the infinitesimal temperature limit: as one raises the temperature, the horizon extends outward toward the IR extremal asymptotic region so that in the full global solution, the UV asymptotics will change. The D-dimensional Lifshitz black holes we are considering in this paper are supported by a scalar field and a U(1) gauge field; their entropy densities, as determined by scaling symmetry, will have the form
\begin{equation}
\hat{s} = cf(\hat{\phi_0})\hat{T}^{\frac{D-2}{z}}
\end{equation}
where hats denote dimensionless quantities, made with appropriate powers of, in our case, the charge density. What we wish to show numerically, as a consequence of the above, is that as one lowers the dimensionless temperature $\hat{T}$ of the black hole while keeping the dimensionless source $\hat{\phi_0}$ fixed, the combination $cf(\hat{\phi_0})$ asymptotes to a fixed value, determined by the value of $\hat{\phi_0}$ and $\xi$:
\begin{equation}\label{app}
\hat{T} \, \rightarrow 0 , \, \hat{\phi_0} \, \text{fixed} \quad \Rightarrow \quad \frac{\hat{s}}{\hat{T}^{\frac{D-2}{z}}} \, \rightarrow \, \text{Constant}
\end{equation}

We form dimensionless parameters with the asymptotic charge density $\rho$, defined by the usual AdS asymptotics as
\begin{equation}
A_t'(r) = \frac{\rho}{r^{D-2}} + \ldots
\end{equation}
Then for a massless scalar in bulk spacetime dimension $D$, and hence with asymptotic falloff
\begin{equation}
\phi(r) \sim \phi_0 + \frac{\phi_1}{r^{D-1}} \, ,
\end{equation} 
we have
\begin{equation}
\hat{s} = \frac{s}{\rho} \, , \quad \hat{T} = \frac{T}{\rho^{\frac{1}{D-2}}} \, , \quad \hat{\phi_0} = \phi_0
\end{equation}

Let us choose the parameters
\begin{equation}
D=4\, , \quad  \alpha=1\, , \quad V_0 = 6 \quad 
\end{equation}
This choice of $\alpha$ gives $z=5$, so that $\hat{s} \sim \hat{T}^{2/5}$. 

We choose to fix the source at $\hat{\phi_0}=1 \pm 10^{-4}$, integrating out to large $r$ such that $U(r)\sim V(r) \sim r^x$, where $x=2 \pm 10^{-5}$. The plot below shows that, indeed, the behavior (\ref{app}) is satisfied down to temperature $\hat{T} \sim 10^{-5}$; further probes of lower temperatures allowed by numerical stability do not deviate from this behavior.

\begin{figure}[h!]
\begin{centering}
  \includegraphics[scale=.75]{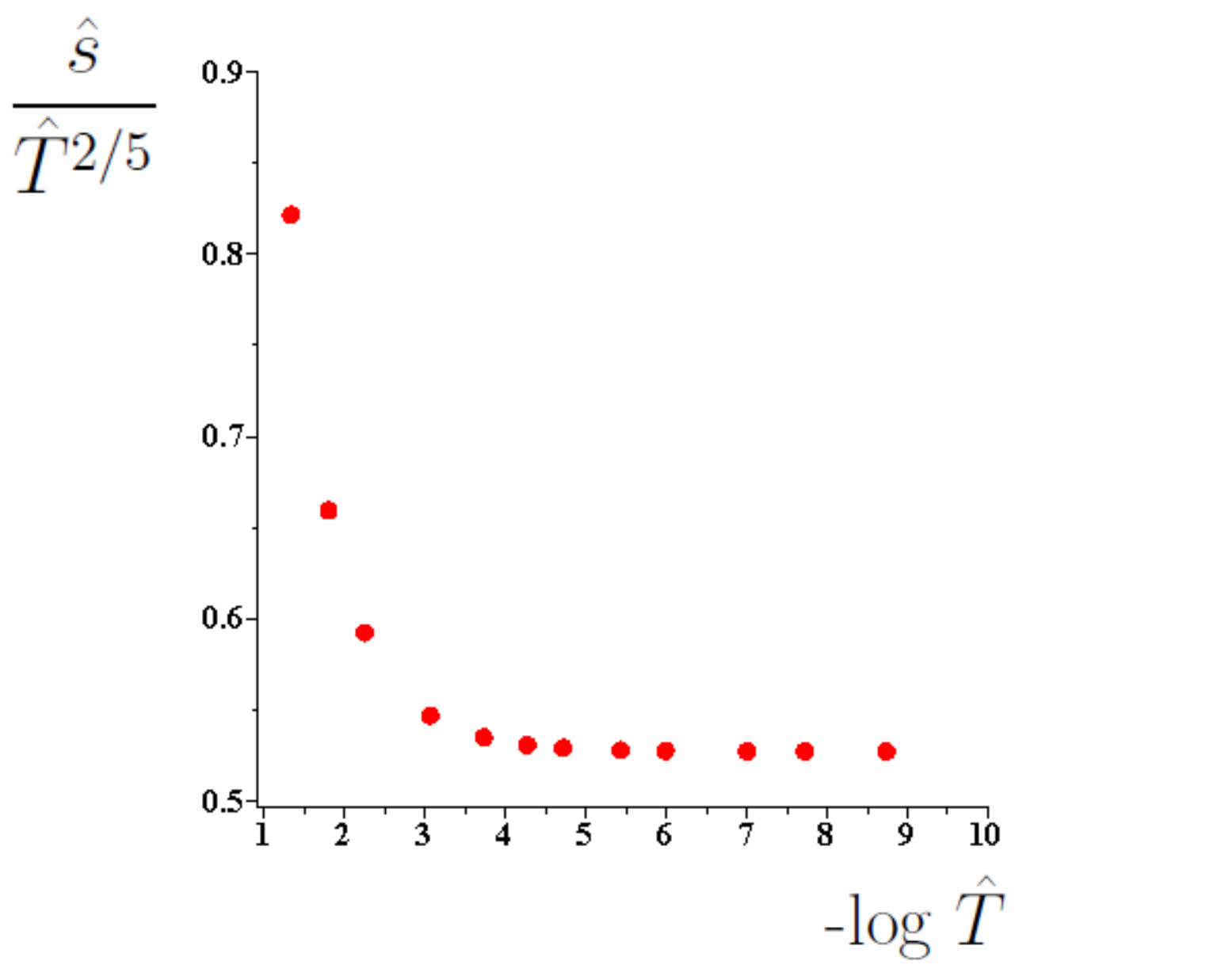}\\
\end{centering}
\caption{Low-temperature power law scaling of entropy density with temperature, including the coefficient. The entropy density displays true Lifshitz power law behavior as the IR asymptotics approach those of the zero temperature case. All quantities are dimensionless.}
\end{figure}

\end{document}